\documentclass[pdflatex]{sn-jnl}

\usepackage{apacite}
\usepackage{url}
\usepackage{hyperref}
\usepackage{graphicx}%
\usepackage{multirow}%
\usepackage{amsmath,amssymb,amsfonts}%
\usepackage{amsthm}%
\usepackage{mathrsfs}%
\usepackage[title]{appendix}%
\usepackage{xcolor}%
\usepackage{textcomp}%
\usepackage{manyfoot}%
\usepackage{booktabs}%
\usepackage{url}
\usepackage{threeparttable}
\usepackage{float}%
\usepackage{booktabs}
\raggedright
\usepackage{graphicx}
\usepackage{caption}
\usepackage{algorithm}%
\usepackage{algorithmicx}%
\usepackage{algpseudocode}%
\usepackage{listings}%
\usepackage{natbib}
\usepackage{tabularray}
\usepackage{etoolbox}
\DeclareMathOperator*{\argmin}{arg\,min}

\let\proglang=\textsf

\theoremstyle{thmstyleone}%

\theoremstyle{thmstyletwo}%

\theoremstyle{thmstylethree}%
\begin{document}


\title[Projections of Minimum Proficiency]{On the Development of Probabilistic Projections of Country-level Progress to the UN SDG Indicator of Minimum Proficiency in Reading and Mathematics}

\author*[1]{\fnm{David} \sur{Kaplan}}\email{david.kaplan@wisc.edu}

\author[2]{\fnm{Nina } \sur{Jude}}\email{jude@ibw.uni-heidelberg.de}

\author[1]{\fnm{Kjorte } \sur{Harra}}\email{harra@wisc.edu}

\author[2]{\fnm{Jonas } \sur{Stampka}}\email{stampka@ibw.uni-heidelberg.de}

\affil*[1]{\orgdiv{Department of Educational Psychology}, \orgname{University of Wisconsin - Madison}, \orgaddress{\street{1025 W. Johnson St.}, \city{Madison}, \postcode{53706}, \state{Wisconsin}, \country{United States}}}

\affil[2]{\orgdiv{Institute for Bildungswissenshaft}, \orgname{Universit\"{a}t Heidelberg}, \orgaddress{\street{Akademiestr. 3}, \city{Heidelberg}, \postcode{69117}, \state{Baden-Württemberg}, \country{Germany}}}

\abstract{As of this writing, there are five years remaining for countries to reach their Sustainable Development Goals deadline of 2030 as agreed to by the member countries of the United Nations. Countries are, therefore,  naturally interested in projections of progress toward these goals. A variety of statistical measures have been used to report on country-level progress toward the goals, but they have not utilized methodologies explicitly designed to obtain optimally predictive measures of \emph{rate of progress} as the foundation for projecting trends.  The focus of this paper is to provide Bayesian probabilistic projections of progress to SDG indicator 4.1.1, attaining minimum proficiency in reading and mathematics, with particular emphasis on competencies among lower secondary school children. Using data from the OECD PISA, as well as indicators drawn from the World Bank, the OECD, UNDP, and UNESCO, we employ a novel combination of Bayesian latent growth curve modeling Bayesian model averaging to obtain optimal estimates of the rate of progress in minimum proficiency percentages and then use those estimate to develop probabilistic projections into the future overall for all countries in the analysis. Four case study countries are also presented to show how the methods can be used for individual country projections.}

\keywords{UN Sustainable Development Goals, Model Uncertainty, Probabilistic Projections}

\maketitle

\section{Introduction}\label{sec1}

\par In 2000, member states of the United Nations adopted the \emph{Millennium Development Goals} (MDGs) \citep{un2015mdg}. Eight goals were advanced, and of relevance to this paper, Goal 2 focused on achieving \emph{universal primary education}. Previously, the 2000 World Education Forum in Dakar, Senegal set out the ambitious goal of [e]liminating gender disparities in primary and secondary education by 2005 and achieving gender equality in education by 2015, with a focus on ensuring that girls enjoy full and equal access to basic education of good quality \citep{EFAGMR2015}.  Under Goal 2 of the MDGs, the specific target (Target 2a) specified that countries ``(e)nsure that, by 2015, children everywhere, boys and girls alike, will be able to complete a full course of primary schooling.  

\par By the end of 2015, considerable progress had been made.  According to \cite{un2015mdg}, substantial progress was realized in (a) increasing enrollment rates, particularly in sub-Saharan Africa; (b) reducing the number of out-of-school children of primary school age; (c) increasing literacy rates world-wide, along with a reduction in the gender gap in reading literacy. Nevertheless, in the developing regions of the world, children in the poorest
households were still four times as
likely to be out of school as those
in the richest households.

\par As the Millennium Development Goals came and went, the UN set about to again stimulate member countries to embrace and achieve the so-called \emph{Sustainable Development Goals (SDGs)}. Again, with a focus on education, UNESCO together with UNICEF, the World Bank, UNFPA, UNDP, UN Women and UNHCR convened the World Education Forum 2015 in Incheon, Republic
of Korea.  Over 1,600 participants from 160 countries, including over 120 ministers, heads and members of delegations, heads of agencies and officials of multilateral and bilateral organizations, and representatives of civil society, the teaching profession, youth and the private sector, adopted the Incheon Declaration for Education 2030, which, at the time, set out a new vision for education for the next fifteen years \citep{unesco2016education}. Specifically, the UN identified equitable, high-quality education, including the achievement of literacy and numeracy by all youth and a substantial proportion of adults, both men and women, as one of its global SDGs to be attained by 2030 \cite{UNSDG}. As of this writing, there are five years remaining to achieve the SDGs generally and the education goals specifically. 

\section{SDG 4: Quality Education}
 
\par The Incheon Declaration recognized that education is the primary driver in meeting many, if not all, of the other SDGs.  The focus of this paper is SDG 4, indicator 4.1.1 which stipulates that the ``[p]roportion of children and young people (a) in grades 2 \& 3; (b) at the end of primary; and (c) at the end of lower secondary [achieve] at least a minimum proficiency level in (i) reading and (ii) mathematics, by sex", and specifically 4.1.1c which concerns the end of lower secondary school.

\par To analyze the potential impact of education policies targeting the SDGs, it is critically important to monitor trends in educational outcomes over time. Indeed, as educational systems around the world face new challenges due to the lingering impact of the COVID-19 pandemic, monitoring trends in educational outcomes could help identify the long-run impact of this unprecedented health crisis on global education. To this end, international large-scale assessment (ILSA) programs such as the Organization for Economic Cooperation and Development (OECD) Program for International Student Assessment (PISA, \citealp{pisa2000FirstResults}  are uniquely situated to provide population-level trend data on literacy and numeracy outcomes \citep{Montoya2022}.

\subsection{Purpose and Significance}

\par The purpose of this paper is to advance an approach for developing  probabilistic projections of the rate of progress that countries are making in reaching or exceeding \emph{minimum proficiency levels} in reading and mathematics.\footnote{From here on, we will simply use the term minimum proficiency rather than minimum proficiency levels.} We rely on data from PISA which assesses in-school 15-year-old students, and hence provides evidence of relevance to SDG indicator 4.1.1c. Following a ``proof-of-concept" paper by \cite{KaplanHuang2021}, we will first derive an estimate of the rate of progress in minimum proficiency across countries using Bayesian latent growth curve modeling.  We will then draw on country-level indicators from PISA as well as data sources from UNESCO and the World Bank to create an initial set of predictors of the rate of progress for boys and girls separately within and across countries. We will then apply \emph{Bayesian model averaging} (BMA), described below, to account for model uncertainty in the selection of the choice of indicators of rate of progress.  We will then use Bayesian model averaged predictors of the rate of progress to generate optimal probabilistic predictions of minimum proficiency in reading and mathematics as defined in PISA.

\par The significance of this paper is three-fold. First, we view the methodologies applied in this paper as contributing to monitoring progress in the internationally agreed-upon education goals of UNESCO, the OECD, the World Bank, the World Education Forum, and others. Second, we view this paper as contributing to sophisticated secondary analyses of large-scale educational assessments.  This paper demonstrates the richness of policy information that can be obtained when using Bayesian probabilistic prediction models to study educational trends at the population level.  Third, this paper is designed to advance Bayesian methodology in education through a novel synthesis of longitudinal models for trend projections with methods that account for uncertainty in said projection models.  

\subsection{Organization}  The organization of this paper is as follows.  In Section 3 we provide an overview of current international trends in reading and mathematics proficiencies along with trend plots for the countries that will be used in this study. Section 4 introduces the method of Bayesian latent growth curve modeling that we will use for estimating rate of progress in achieving minimum proficiency.  In Section 5, we provide a brief review of Bayesian decision theory which motivates the method we use to obtain optimal predictions of growth.  In Section 6 we introduce BMA which is the method we use to obtain optimal estimates of growth while taking into account model uncertainty.  Section 7 discusses the data sources, selection of indicators, and methods for handling missing data, along with the workflow for this paper.  We introduce the main data source used to assess change over time in the country-level percentages of students achieving minimum proficiency in reading and mathematics, namely PISA\footnote{Unless otherwise stated, for simplicity we will use the phrase ``achieving minimum proficiency" where it is understood that we are referring to both boys and girls and for both reading and mathematics}. Section 8 presents the growth curve modeling and the model averaging results. 
 Section 9 presents the overall projections across the countries used in this analysis as well as our case studies.  Section 10 concludes.

\section{Overview of Trends in Reading and Mathematics Literacy}
\par A major substantive focus of this paper is educational equity, and in particular the impact of COVID-19 on exacerbating inequities in educational outcomes. Educational equity at the system level can be defined as schools providing equal learning opportunities for all students \citep{OECD2018}.  Equity can be seen as a fundamental goal of education and a guiding principle in education policy that is accepted worldwide \citep{UIS2018}. In an ideal world, students' characteristics, such as family background or gender, should be unrelated to  educational outcomes such as literacy in different domains. However, countries clearly differ in their level of equity already attained as well as in the national policies they implement to reach agreed-upon goals. 

\par Policy measures promoting equity can be directed at the individual, the institution, or the system level \citep{LevinB2003}. Educational systems that are less stratified and provide more flexible educational pathways tend to show more equity in educational outcomes \citep{OECD2008}. Opposite effects can be seen in systems where parents (or students) are given free school choice. This often leads to increased differences in the social composition of schools, which is usually related to aspects of equity. For example, gender gaps can be lower in high-socioeconomic status schools \citep{BorgonoviEtAl2018}.  


\par In the case of gender equity as a specific goal, significant progress has been made and the global impact of this progress can be seen in current education statistics.  For example, with regard to school enrollment rates, parity was achieved globally, on average, up to upper-secondary education \citep{UNESCO2016}. Still, gender gaps in literacy exist in all countries where data are available; although girls usually outperform boys in reading, the opposite is the case for mathematics and science. While the actual gender gap in students' literacy could be considered comparatively small in many of the OECD countries, these effects do not vanish, but rather increase over time.  For example, results from the OECD Program for the International Assessment of Adult Competencies (PIAAC), which focuses on adult populations, showed that competence in numeracy was significantly higher for men than women in all participating countries \citep{Encinas-Martin2023}. 

\par While graduation rates in tertiary education are now comparable between men and women in most OECD countries \citep{Schleicher2008}, gender differences related to the STEM subjects remain high. The result is a disadvantage for women in accessing this rapidly developing share of the labor market \citep{UNESCO2016}.

\par We have so far described the state-of-affairs prior to the COVID-19 pandemic.  This unprecedented health crisis has the potential of exacerbating the gender disparities in education between men and women. Current research already points to significant impacts of the pandemic on educational outcomes that will probably be seen in the years to come \citep{naep,WorldBank2022LearningPoverty}.  Connected to the impacts of the pandemic on educational outcomes, data collected by the United Nations shows that the impact of the pandemic threatens to deepen gender poverty gaps \citep{UNWomenUNDP2020}. However, effects might be country specific, as the TIMSS 2023 results show that, in many countries, the performance in mathematics and science remained stable or improved compared to 2019, while others experienced learning losses particularly in mathematics \citep{vonDavier2024TIMSS}.

\par Clearly, challenges remain in identifying the origins of gender differences in major literacy domains. Policymakers need to be aware of the long-term influence of gender gaps existing at school-age \citep{BorgonoviEtAl2017}, and especially now in tracking gender gaps resulting from the pandemic.  Consequently, tracking the development of gender equity over time is crucial to international educational policy. 

\par We will focus on country-level longitudinal outcomes in minimum proficiency using data from PISA 2009 to PISA 2022.  Although longer time points are available, it was decided to only include countries with complete data on the reading and mathematics outcomes so that predictive analyses were not dependent on the imputation of excessive amounts of missing data. The list of these countries is shown in Table~\ref{tab:countries}.

\begin{table}[h!]
\centering
\caption{Fifty-three countries and economies with complete reading and mathematics assessment data from 2009 to 2022.}
\label{tab:countries}
\begin{tabular}{rrrrr} 
\toprule
Albania & Argentina & Australia & Austria & Belgium \\
Brazil & Bulgaria & Canada & Chile & Colombia \\
Croatia & Czech Republic & Denmark & Estonia & Finland \\
France & Germany & Greece & Hong Kong–China & Hungary \\
Iceland & Indonesia & Ireland & Israel & Italy \\
Japan & Jordan & Kazakhstan & Korea & Latvia \\
Lithuania & Macao–China & Mexico & Montenegro & Netherlands \\
New Zealand & Norway & Peru & Poland & Portugal \\
Qatar & Romania & Singapore & Slovak Republic & Slovenia \\
Spain & Sweden & Switzerland & Thailand & Turkey \\
United Kingdom & United States & Uruguay &  &  \\ 
\bottomrule
\multicolumn{5}{l}{\textit{Note:} Luxembourg did not have data for PISA 2022, and the Russian Federation was} \\
\multicolumn{5}{l}{excluded from the PISA 2022 assessment due to the war in Ukraine.}
\end{tabular}
\end{table}

\par Two points are worth noting regarding the countries shown in Table \ref{tab:countries}.  First, these are countries who are either members of the OECD or partner countries that are participating in PISA, and although there is a range of economic development covered by these countries, all of these countries are considered economic democracies as defined by the OECD.  Second, it is important to note that much of the global south countries are not represented in this list; countries in sub-Saharan Africa, the Indian sub-continent, and China.  This is relevant insofar as the vast majority of the world's children are located in China and India.  The major reason for the exclusion of these countries is simply the lack of data, and where there are data sources at regional levels, the ability to provide reliable links to PISA remains an open methodological question.  This issue will be taken up in the Conclusions section.

\begin{figure}[h]
  \centering
  {\includegraphics[scale=.18]{{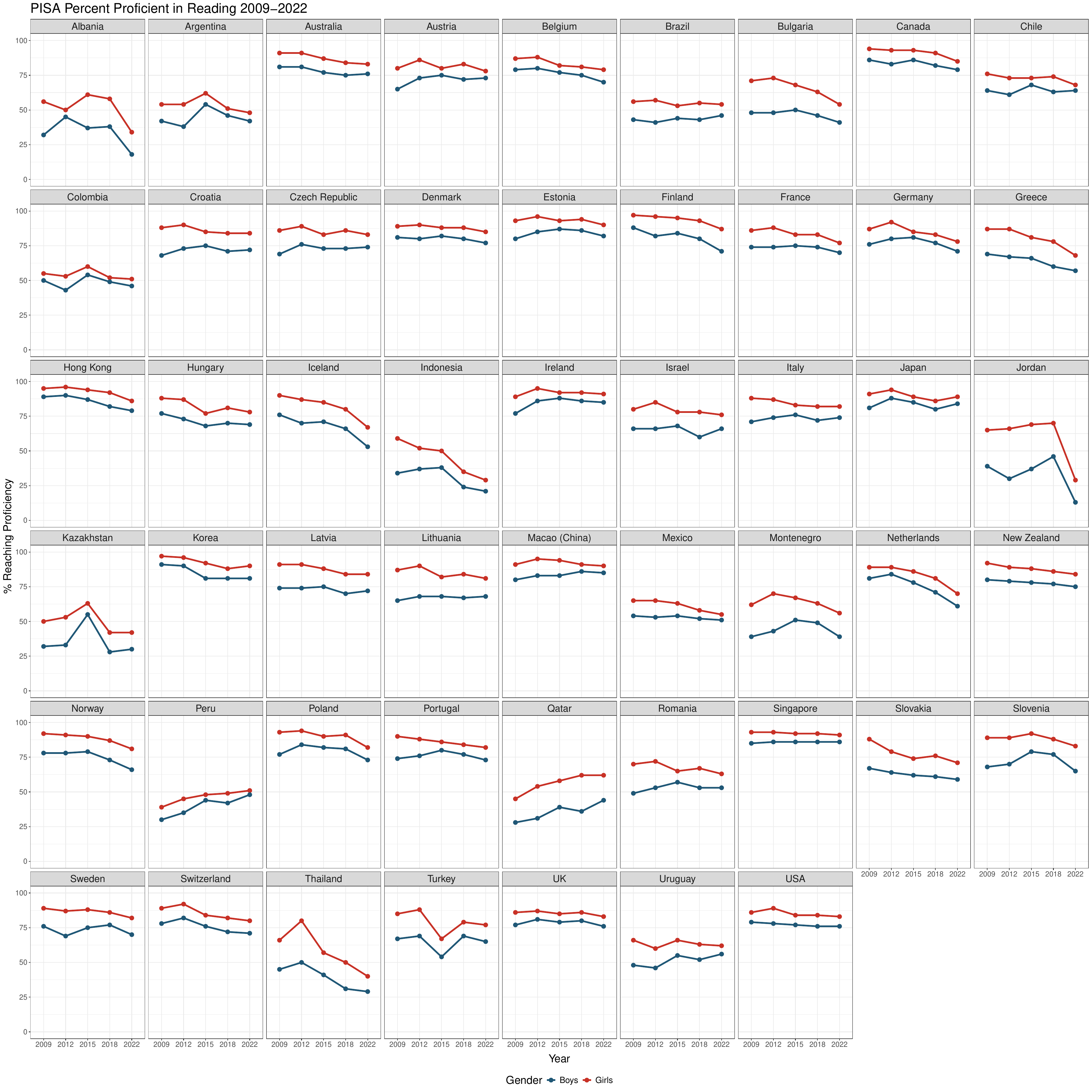}} 
   \caption{Trend lines for percentage of students at the minimum proficiency level in reading from 2009 to 2022.  The red line is for the girls, the blue line is for boys. A downward slope indicates a decrease in the percentage attaining the minimum proficiency level in reading.}\label{fig:pisareading}}
\end{figure}

\begin{figure}[H]
    \centering
  {\includegraphics[scale=.18]{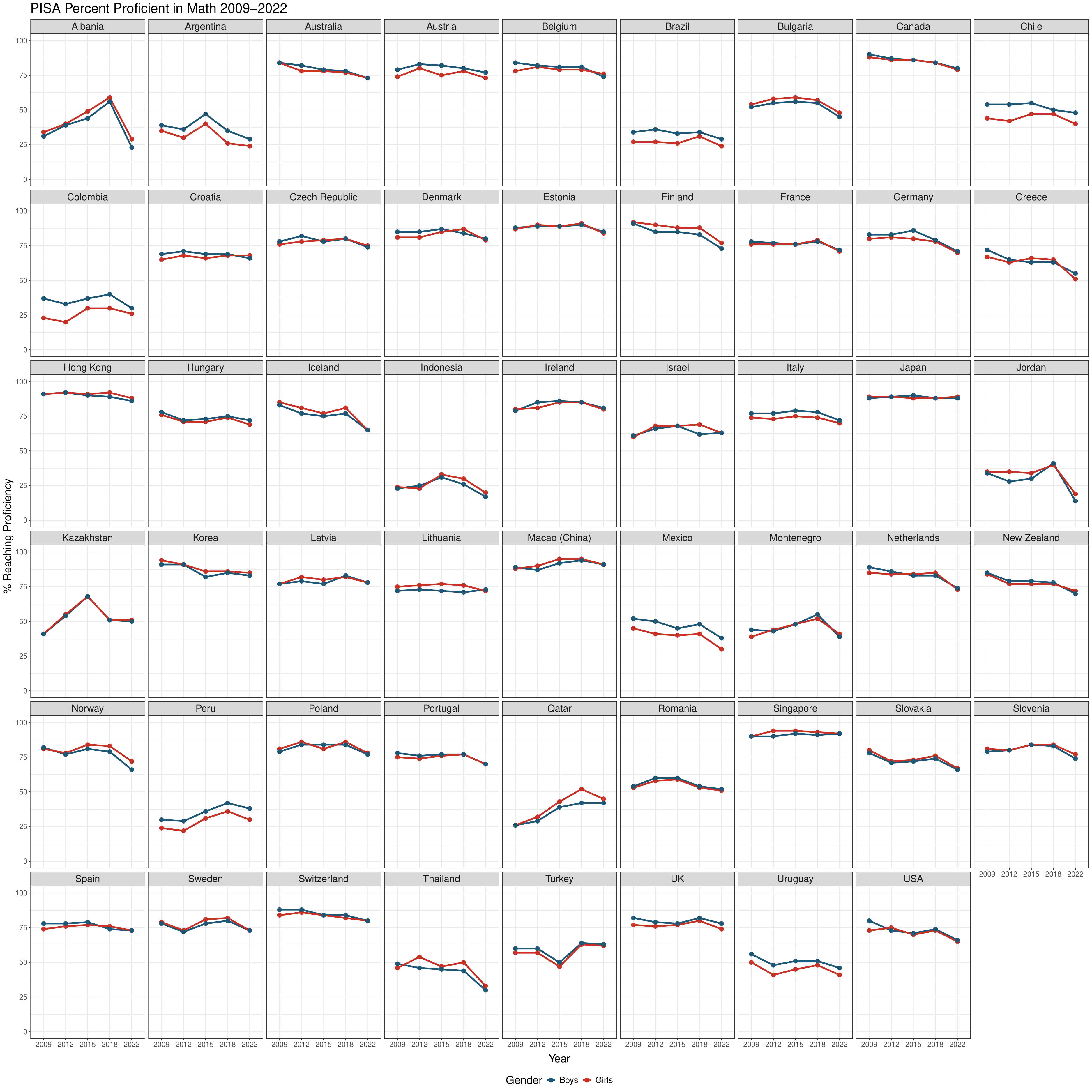} 
  \caption{Trend lines for percentage of students at the minimum proficiency level in mathematics from 2009 to 2022.  The red line is for the girls, the blue line is for boys. A downward slope indicates a decrease in the percentage attaining the minimum proficiency level in mathematics.}\label{fig:pisamath}}
  \end{figure}

The longitudinal outcome will be the percentage of students at the country level, male and female, who have reached the PISA-defined minimum proficiency level across five cycles of PISA.  For reading, the PISA minimum proficiency lower cut score is 407 defining \emph{Level 2} of the PISA proficiency scale.  For mathematics, the lower cut score for minimum proficiency is 420.  Figures (\ref{fig:pisareading}) and (\ref{fig:pisamath}) show the changes over time in these percentages. 


\clearpage

\par Inspection of Figures (\ref{fig:pisareading}) and (\ref{fig:pisamath}) reveal some interesting features. Figure (\ref{fig:pisareading}) shows the well-documented differences between the percentage of girls and boys at the minimum proficiency in reading.  That said, the trend over time for boys and girls, while somewhat variable across countries, are basically parallel to each other within countries with some countries showing gains in the  percentage at minimum proficiency and others declining.  For mathematics proficiency shown in Figure (\ref{fig:pisamath}), the advantage favoring boys is not as clear, and it appears that again, although there is some variability in the trend across countries, very little difference is observed between boys and girls over years 2009 to 2022.  Finally, for both reading and mathematics, and for both boys and girls, the shape of the trend lines is mostly linear for the countries in this analysis with a general downward trend in most countries.  Some countries exhibit a steep decline in 2022, which could be due to the impact of the pandemic, but definite causal conclusions cannot be drawn.

\section{Bayesian Growth Curve Modeling}
With the empirical trajectories in hand, the next step is to estimate the average starting level and rate of progress in the percent at or above minimum proficiency over the cycles of PISA. We frame our estimation of these parameters as a Bayesian hierarchical latent variable model \citep{kaplanbayesbook2}. 

Let the \emph{within-country} model be written as follows
\begin{equation}\label{L1GCM}
\boldsymbol{y}_{i} = \boldsymbol{\Lambda}\,\boldsymbol{\eta}_{i} +\boldsymbol\epsilon_{y(i)}
\end{equation}
where $\boldsymbol{y}_{i}$ is a $T \times 1$ vector of $T$ waves of measurement for country $i$ ($i = 1,\ldots,n$); $\boldsymbol{\Lambda}$ be a $T \times Q$ matrix of fixed constants that for simplicity are assumed to be the same for all countries.  These terms can be specified to represent linear, quadratic, or even higher order shapes of the trajectories and also serve to parameterize the growth model as a structural equation model \citep{willettsayer94, muthen1997latent, BollenCurran06, GrimmRamEstabrook2017}.  Further, let $\boldsymbol{\eta}_{i}$ be a $Q \times 1$ vector of random growth parameters for country $i$, where the random growth parameters can be considered as latent variables. In our study, $Q=2$, representing the intercept and linear growth component for each country; and  $\boldsymbol\epsilon_{y(i)}$ is a $T \times 1$ vector of residuals,
with diagonal covariance matrix
\begin{equation}\label{sigmay}
\boldsymbol\Sigma_{y_{i}} = 
\begin{bmatrix}
    \sigma^2_{y_{1(i)}} & &    \\
    & \sigma^2_{y_{2(i)}} & \\
    & & \ddots &\\
    & & & \sigma^2_{y_{(T-1)(i)}}\\
    & & & & \sigma^2_{y_{T(i)}}\\
\end{bmatrix}.
\end{equation}

\par The \emph{between-country model} allows the random growth parameters to be related to a set of time-invariant predictors.  The level-2 model can be written as
\begin{equation}\label{L2GCM}
\boldsymbol{\eta}_{i} = \boldsymbol{\Gamma}\boldsymbol{x}_{i} + \boldsymbol\epsilon_{\eta(i)},
\end{equation}
where $\boldsymbol{\Gamma}$ is a $Q \times P$ vector of regression coefficients associated with the time-invariant predictors, $\boldsymbol{x}_{i}$ is a $P \times 1$ vector of time-invariant predictors, and 
$\boldsymbol\epsilon_{\eta(i)}$ is a $Q \times 1$ vector of residuals with symmetric covariance matrix
\begin{equation}\label{EtaCov}
\boldsymbol\Sigma_{\eta_{i}} = 
\begin{bmatrix}
    \sigma^2_{\eta_0}      & sym.  \\ 
    \sigma_{\eta_1\eta_0}& \sigma^2_{\eta_1} \\

\end{bmatrix},
\end{equation}
allowing for the random growth parameters (i.e., intercept and slope) to be correlated conditional on the predictors as show in Eq. (\ref{EtaCov}).

\subsection{Bayesian Hierarchical Growth Curve Model}
\par A Bayesian hierarchical specification of the growth curve model in Eqs. (1 - 6) can be written as
\begin{subequations}\label{HBGCM}
\begin{align}
\boldsymbol{y}_{i} &\sim N(\boldsymbol{\Lambda}\boldsymbol{\eta}_{i}, \boldsymbol\Sigma_{y(i)}),\label{HBGCM1}\\
\boldsymbol{\eta}_{i} &\sim N(\boldsymbol{\Gamma}\boldsymbol{x}_{i},\boldsymbol\Sigma_{\eta_{i}}),\label{HBGCM2}\\
\boldsymbol{\Gamma} &\sim N(\boldsymbol{\Omega},\boldsymbol\Sigma_{\Gamma}),\label{HBGCM3}
\end{align}
\end{subequations}
where $\boldsymbol\Omega$ and $\boldsymbol\Sigma_\Gamma$ are fixed and known parameters.  Prior distributions on the residual covariance matrices are assumed to be inverse-Wishart (IW). For the current data, let $\sigma^{0}_{y_{i}}$ indicate the standard deviation of level-1 residual, which is the square root of the diagonal components in Eq. (\ref{sigmay}). The prior distributions for the variance parameters are specified as follows:

\begin{subequations}
\begin{align}
\sigma^{0}_{y_{i}} &\sim \mbox{half-Cauchy}(\mu_y, \sigma_y),\\
\boldsymbol{\Sigma}_{\eta_{i}} &\sim \mbox{IW}(\textbf{R}_\eta, \nu_\eta),
\end{align}
\end{subequations}
but of course other prior specifications are possible \citep[see][]{kaplanbayesbook2}.

\subsection{Flexible Models for rate of progress}
\par An important flexibility in growth curve modeling allows for the estimation of non-linear trajectories using so-called \emph{latent basis} methods. This specification requires that some of the time points be fixed to constant values for model identification purposes while allowing the remaining time points to be estimated from the data directly. Latent basis modeling yields data-based estimates of the time points and often provides better fit of the model to the empirical trajectories of change over time compared to forcing, say, a linear trend on the data. Within the framework of latent basis methods, the rate parameter $\pi_{1i}$ is best conceived of as a \emph{shape} parameter, but we will continue to refer to this as a rate of progress parameter. These, and other extensions, are discussed in \cite{BollenCurran06} and \cite{GrimmRamEstabrook2017}.  
\par For this paper, we will examine differences in predictive performance across three latent growth curve models: (a) a simple linear growth curve model with fixed and equally spaced time points, (b) a latent basis model with that last time point (2022) freed to reflect a possible decrease in scores due to the COVID-19 pandemic, and (c) a latent basis model in which the basis functions from 2015-2022 are freed, reflecting the idea that the downward trend had started prior to 2022 \citep[see also][]{KaplanHarraStampkaJude2025}

For this paper, the inter-country model for the starting minimum proficiency percentage $\pi_{0i}$ and rate of progress $\pi_{1i}$ respectively, can be written as
\begin{equation}\label{l2int}
\pi_{0i} = \beta_{00} + \sum_{q = 1}^{Q}\beta_{0q}x_{qi} + \epsilon_{\pi_{0i}}, \qquad {(i = 1,\ldots,n)}
\end{equation}
and
\begin{equation}\label{l2slp}
\pi_{1i} = \beta_{10} + \sum_{q = 1}^{Q}\beta_{1q}x_{qi} + \epsilon_{\pi_{1i}}, \qquad {(i = 1,\ldots,n)}
\end{equation}
where $x_{qi}$ are values on $Q$ predictors for country $i$, $\beta_{00}$, $\beta_{01}$, $\beta_{10}$, and $\beta_{1q}$ are the intercept and slopes associated with the time-invariant predictors of the starting minimum proficiency percentage and rate of progress, and  $\epsilon_{\pi_{0i}}$ and $\epsilon_{\pi_{1i}}$ are errors. An example of a time-invariant predictor might be the type of political system of a country.\footnote{Of course, this assumes that the political system of a country is relatively stable over time.} For this paper, we will develop a set of distinct models for predicting $\pi_{0}$ and $\pi_{1}$.\footnote{Note that Equations (\ref{l2int}) and (\ref{l2slp}) imply that the same predictors are being used for the starting minimum proficiency percentage and rate of progress, and although that will be the case for this paper, it is not necessary, and different predictors for these parameters can be specified.} 

\section{The Decision-Theoretic Framework for Predictive Modeling}
\par Foundational to our goal of developing an \emph{optimally predictive} model to be used to forecast the rate of progress in minimum proficiency is \emph{Bayesian decision theory}.  Bayesian decision theory structures the problem of predictive modeling in the context of minimizing \textit{expected loss} -- that is, the penalty that  results from using a particular model to predict future observations.  The less the expected loss, the better the model is at predictive performance in comparison to other models. Bayesian decision theory (see e.g., \cite{Good1952,lindley1991making,berger2013statistical}) provides a natural and intuitive foundation for developing and evaluating Bayesian predictive models.  

\subsection{Fixing Notation and Concepts}
\par  An outline of our Bayesian decision-theoretic framework was given in \cite{KaplanHarraStampkaJude2025} and is reproduced here.  
Let $D = \{y_{ci}, x_i, z_{ci}\}_{i=1}^n$ be a set of data that is assumed to be fixed in the Bayesian sense where one conditions on observable and fixed data, and where $y_{ti}$ is the outcome of interest at cycle $t$ for country $i$, $x_i$ is a (possibly vector-valued) set of time-invariant predictors, and $z_{ti}$ is a set of time varying-predictors for country $i$.  Additionally, let ($\tilde{y}, \tilde{x},\tilde{z}$) be a future observation of the outcome of interest and the set of predictors, respectively.  Finally, let $\mathcal{M}=\{{M}_k\}_{k=1}^K$ represent a set of individual models that are specified to yield predictions of the rate of progress in minimum proficiency, $\pi_1$ and let $M_k$ represent a specific model for the rate of progress. Each $M_k$ will eventually be a member in the ensemble $\mathcal{M}$.

\par The features of Bayesian decision theory that we adopt in this paper have been described elsewhere by \cite{bernardosmith2000} and \cite{VehtariOjanen2012} among many others. These elements consist of (a) an unknown state of the world denoted as $\omega \in \Omega$, (b) an action $a \in \mathcal{A}$, where $\mathcal{A}$ is the action space, (c) a loss function $L(a,\omega): \mathcal{A} \times \Omega \rightarrow \mathbb{R}$ that rewards an action $a$ when the state of the world is realized as $\omega$, and (d) $p(\omega|D)$ representing one's current belief about the state of world conditional on observing the data, $D$.

\par To provide a context for these ideas, and in anticipation of our application to minimum proficiency for indicator 4.1.1c, consider the problem of predicting a future percentage of boys and girls reaching or exceeding minimum proficiency. Following \cite{bernardosmith2000}, \cite{lindley1991making}, \cite{VehtariOjanen2012} and \cite{berger2013statistical} and the notation given previously for country $i$ at time $t$, (a) the states of the world correspond to the future minimum proficiency percentages from future cycles of PISA, that is, $\tilde{y} \in \mathcal{Y}$, (b) the action $a \in \mathcal{A}$ is the actual prediction of those future observations based on using an optimized prediction of the rate of progress $\pi_{1}$, (c) the loss function $L(a,\tilde{y})$ defines the loss attached to the prediction, and (d) a posterior predictive distribution, $p(\tilde{y}|D, M_*)$, that encodes our belief about the rate of progress in minimum proficiency percentages conditional on the data, $D$.

\subsection{Loss Functions for Evaluating Predictions}
\par The intent of predictive modeling is to minimize the loss associated with taking an action $a$ among a set of actions included in an action space $\mathcal{A}$.  A number of loss functions exist, but common loss functions rely on the negative \textit{quadratic loss} function
\begin{equation}
L(a,\tilde{y} ) = (\tilde{y} - a)^2.
\end{equation}
The optimal action $a^*$ is the one that minimizes the \textit{posterior expected loss}, written as 
\begin{equation}
a^* = \argmin_{a \in \mathcal{A}} \int_{\Omega}L(\omega, a)p(\omega|D)d\omega.
\end{equation}
The objective of predictive modeling, therefore, is to take an action $a$ that minimizes the loss $L$ when the future observation is $\tilde{y}$. \cite{ClydeIversen2013} showed that the optimal decision $a^*$ obtains when $a^* = E(\tilde{y}|D)$, which is the posterior predictive mean of $\tilde{y}$ given the data $D$.  Assuming that the correct data generating model exists and is among the set of models under consideration, this can be expressed as
\begin{equation}\label{bmapostmean}
E(\tilde{y}|D) = \sum_{k=1}^{K}E(\tilde{y}|M_k,D)p(M_k|D)=\sum_{k=1}^{K}p(M_k|D)\hat{\tilde{y}}_{M_k},
\end{equation}
where $\hat{\tilde{y}}_{M_k}$ is the posterior predictive mean of $\tilde{y}$ under $M_k$  and $p(M_k|D)$ is the \emph{posterior model probability} (PMP) associated with model $k$.  The PMP can be written as
\begin{equation}\label{PMP}
p(M_k|D) = \frac{p(D|M_k)p(M_k)}{\sum_{l=1}^Kp(D|M_l)p(M_l)}, \qquad{l \ne k}
\end{equation}
where $p(M_k)$ the prior probability for model $k$. Equations (\ref{bmapostmean}) - (\ref{PMP}) define BMA.\footnote{The assumption that the true model is in the space of models under investigation is referred to as the $\mathcal{M}$-closed framework \citep{bernardosmith2000}.  This a rather difficult assumption to warrant in the education and the social sciences generally.  A relaxation of this assumption is referred to as the $\mathcal{M}$-open setting \citep{bernardosmith2000}. A recent paper with applications to PISA 2018 by \cite{Kaplan2021Psychometrika} compared BMA with \emph{Bayesian stacking} (e.g.\cite{Breiman1996}, which is suited for the $\mathcal{M}$-open settings and found BMA to provide slightly better predictive performance.  Thus, in following previous studies that have applied BMA to to growth regression in economics (e.g. \cite{fernandezleysteel01b} we will use BMA for this study.}    By default, a uniform prior probability mass is placed over the model space and non-informative unit information priors \citep{kasswasserman} are used for the parameters of each model, but software programs such as \proglang{BMS} \citep{BMS} allow for other prior distribution choices over both the parameters and the model space.  We will examine the sensitivity of our BMA analysis to the choice of priors in our empirical analysis below.  

\section{Bayesian Model Averaging}
\par When addressing the problem of developing probabilistic projection models, it is common to specify and estimate a set of different projection models and to use various model selection methods such as Akaike's information criterion (AIC) \citep{akaike85,akaike87} or the Bayesian information criterion (BIC) \citep{Schwarz78, kassraftery} to choose a final model to report.  The difficulty with model selection methods is that the analyst often proceeds as though the final selected projection model was the one considered in advance, and thus the uncertainty in the model selection process is ignored.  More to the point, the selection of a particular model from a universe of possible models can be characterized as a problem of uncertainty.  The problem of model uncertainty has been succinctly characterized by Hoeting and her colleagues \citep{hoeting99}: 

\begin{quote}Standard statistical practice ignores model uncertainty.  Data analysts typically select a model from some class of models and then proceed as if the selected model had generated the data.  This approach ignores the uncertainty in model selection, leading to over-confident inferences and decisions that are more risky than one thinks they are. (p. 382) \end{quote}

\par To address the problem of model uncertainty, we propose to use BMA \citep{leamer,madiganraftery94,raftery97,hoeting99}. The key idea of BMA (described in more detail below) is to recognize that selecting a single model out of a class of possible models that could have been selected effectively ignores the uncertainty inherent in model choice.  Model averaging diminishes the uncertainty in the choice by taking a weighted average over a selected set of models, with weights that account for the quality of the model. These weights are the posterior model probabilities (PMPs).
\par Of importance to this paper is that theory and applications of BMA have shown that it provides better predictive performance to that of any single model based on a class of scoring rules used in probabilistic forecasting  \citep{raftery97}. Below, we will employ two specific scoring rules that we will use to evaluate the predictive performance of our projection models.

\par Bayesian model averaging has had a long history of theoretical developments and practical applications.  Early work by \cite{leamer} laid the foundation for BMA. Fundamental theoretical work on BMA was conducted in the mid-1990s by Madigan and his colleagues (e.g., \cite{madiganraftery94,raftery97,hoeting99}).  Additional theoretical work was conducted by \cite{clyde99}.  \cite{draper95} has discussed how model uncertainty can arise even in experimental designs, and \cite{kassraftery} provide a review of BMA and the consequences of ignoring model uncertainty. A more general review of the problem of model uncertainty can be found in \cite{clydegeorge}.  

\par In addition to theoretical developments, BMA has been applied to a wide variety of content domains.  A perusal of the extant literature shows BMA applied to economics (e.g. \cite{fernandezleysteel01b}),  bioinformatics of gene express (e.g. \cite{yeungbumgarnerraftery}), weather forecasting (e.g., \cite{sloughteretal}), and causal inference within propensity score analysis \citep{kaplanchenbma} to name just a few.  An extension of BMA to structural equation modeling with applications to education can be found in \cite{kaplanleebmasem}, and, of relevance to this paper, a general review of BMA in the context of cross-sectional analyses of large-scale educational assessment data can be found in \cite{KaplanLee2018}. A novel implementation of BMA to multiple imputation was proposed by \cite{miBMA}.

\subsection{Technical Background of BMA}
\par The aim of this paper is to provide a set of methods and practices that can yield probabilistic projections of country-level progress in meeting or exceeding minimum proficiency in reading and mathematics.  Following \cite{madiganraftery94}, we denote a future outcome of interest (in our case, country-level mathematics or reading minimum proficiency) as $\Upsilon$.  Next, consider a set of competing models $\{{M}_k\}_{k=1}^K$ that are not necessarily nested.  The posterior distribution of $\Upsilon$ given data $y$ can be written as
\begin{equation}\label{sumdelta}
p(\Upsilon|y) = \sum_{k=1}^K p(\Upsilon|M_k)p(M_k|y),
\end{equation}
where $p(M_k|y)$ is the posterior probability of model $M_k$ written as
\begin{equation}\label{postmodel}
p(M_k|y) = \frac{p(y|M_k)p(M_k)}{\sum_{l=1}^Kp(y|M_l)p(M_l)}, \qquad{l \ne k}.
\end{equation}
The interesting feature of Equation (\ref{postmodel}) is that $p(M_k|y)$ will likely be different for different models.  The term $p(y|M_k)$ can be expressed as an integrated likelihood
\begin{equation}\label{intlike}
p(y|M_k) = \int p(y|\theta_k,M_k)p(\theta_k|M_k)d\theta_k,
\end{equation}
where $p(\theta_k|M_k)$ is the prior distribution of $\theta_k$ under model $M_k$ \citep{raftery97}. Bayesian model averaging thus provides an approach for combining models specified by researchers, or perhaps elicited by relevant stakeholders.  

\subsection{Computational considerations for BMA}
The software program we will use in this paper is \proglang{BMS} \citep{BMS}, which is part of the \proglang{R} programming environment \citep{cran}. The default algorithm within \proglang{BMS} is referred to as the birth-death (BD algorithm).  The BD algorithm a Metropolis-Hastings-type Markov Chain Monte Carlo (MCMC) method designed to explore the vast space of possible regression models. In BMA, each model corresponds to a specific subset of covariates. Since the total number of models grows exponentially with the number of covariates (e.g., $2^K$ for $K$ predictors), the BD algorithm samples models rather than enumerating all of them.\footnote{This does not count interactions in which case the model space is $2^{K + \binom{K}{2}}$. Interactions can be handled in the \proglang{BMS} program but we omit them in this study.}
It does this by proposing local moves: a \emph{birth} move adds one variable to the current model, while a \emph{death} removes one variable. These moves ensure that the Markov chain eventually approximates the posterior distribution over the model space.

\par At each step, the algorithm randomly chooses whether to attempt a birth or a death move (with roughly equal probability). It then proposes a new model by either adding a new covariate (birth) or removing one (death), and calculates the posterior model probability (PMP) of the proposed model. The move is accepted with a probability that depends on the ratio of the posterior probabilities of the new and current models, adjusted for the proposal probabilities. This acceptance mechanism ensures that models with higher posterior probabilities are more frequently visited, allowing the empirical frequency of visits to approximate the true posterior model probabilities.

\par The BD algorithm is particularly efficient in large model spaces because it focuses sampling on high-probability regions while still allowing occasional exploration of lower-probability models. Over many iterations, this results in a posterior distribution over models and predictors that can be summarized in terms of posterior inclusion probabilities (PIPs), predictive densities, and model rankings. The algorithm naturally favors models with strong support in the data, and if the true data generating model is in the space of models under consideration, then the BD algorithm will asymptotically concentrate its sampling around the true model, if the true model is not in the space, on models with highest posterior mass (see also Footnote 4).

\subsection{Choosing Parameter and Model Priors in BMA}
\par A required step when implementing BMA is to choose both parameter and model priors \citep{fernandezleysteel01a, liang2008mixtures, eicher2011default, feldkircher2009benchmark}.  In BMA, parameter priors are variations of Zellner's $g$-prior \citep{ZellnerGPrior}. Specifically, Zellner  introduced a natural-conjugate Normal-Gamma $g$-prior for regression coefficients $\beta$ under the normal linear regression for model, written as,
\begin{equation}
y_i = x'_i\beta + \varepsilon,
\end{equation}
where $\varepsilon$ is i.i.d. $N(0, \phi^{-1})$. For a give model, say $M_k$, Zellner's $g$-prior can be written as 
\begin{equation}
\beta_k|\sigma^2,M_k,g \sim N\bigg(0, \sigma^2g\big(x_k'x_k\big)^{-1}\bigg). 
\end{equation}
\par Implementing the $g$-prior requires a choice between \emph{fixed} forms of the $g$-prior and \emph{flexible} forms of the $g$-prior \cite{feldkircher2009benchmark}. Fixed forms of the $g$-prior require that the $g$ be set to a particular value, typically related to the sample size, the number of predictors or both.  These include (a) the \emph{Unit Information Prior} (UIP, $g$ = $N$); (b) the \textit{Risk Inflation Criterion Prior} (RIC, $g = Q^2$); (c) the \emph{Benchmark risk inflation criterion} (BRIC, $g = \mbox{max}(N, Q^2)$; and (d) the \emph{Hannan and Quinn Priors} (HQ, $g = \mbox{log}(N)^3$).

\par As pointed out by \cite{feldkircher2009benchmark}, a problem with fixed  priors is that they are \emph{dengenerate} in the sense that all the prior sets all the probability to one point.  Instead of only employing degenerate priors, \cite{liang2008mixtures} introduced the \emph{hyper-$g$-prior} that ``let the data choose'' \citep[see also][]{feldkircher2009benchmark}, thus reducing the sensitivity of the prior choice of the fixed $g$-prior on the posterior mass. This family of priors was proposed for data-dependent shrinkage. Following \cite{feldkircher2009benchmark}, the hyper-$g$ prior is a Beta prior on the shrinkage factor $\frac{g}{1+g}$, that is $p\big(\frac{g}{1+g}\big) \sim Beta(1, \frac{\alpha}{2} - 1)$, with $E\big(\frac{g}{1+g}\big) = \frac{2}{\alpha}$. Instead of eliciting $g$ directly,the hyper-$g$ prior requires the elicitation of the hyperparameter $\alpha \in (2, \infty)$. As $\alpha$ approaches 2, the prior distribution on the shrinkage factor $\frac{g}{1+g}$ will be close to 1; while for $\alpha$ = 4, the prior distribution on the shrinkage factor will be uniform distributed. In the context of noisy data, the hyper-$g$ prior will distribute posterior model probabilities more uniformly across the model space.  In the case of low noise in the data, the hyper-$g$ prior will be concentrated on a few models, and perhaps even more concentrated than in the fixed prior case with large $g$ \citep{feldkircher2009benchmark}.

\par For this paper, three model priors are investigated and are available in the \textsf{BMS} program: (a) the uniform model prior, (b) the binomial model prior, and (c) the beta-binomial model prior.  The uniform model prior is a common default prior for Bayesian model averaging.  Specifically, if there are $Q$ predictors, then the prior on the space of models is $2^{-Q}$. The difficulty with the uniform model prior was pointed out by \cite{BMS} who noted that the uniform model prior implies that the expected model size is $\sum_{q=0}^Q{\binom{Q}{q}q2^{-Q}} = Q/2$. However, the distribution of model sizes is not even. The result is that the uniform model prior actually places more mass on intermediate size models.  A demonstration of the impact of this problem is given in \cite{BMS}.

\par To address the problem with the uniform model prior, \cite{BMS} proposed placing a fixed inclusion probability on each predictor in the model, denoted as $\theta$.  Then, for model $k$, the prior probability for a model of size $q$ is $p(M_k) = \theta^{q_k}(1-\theta)^{Q-q_m}$.  Notice that the expected model size, say $\bar{m}$, is $Q\theta$, and thus the analysts prior expected model size is $\bar{m}$. Moreover, if $\theta = .5$, then the binomial model prior reduces to the uniform model prior.  In practice, this suggests that choosing $\theta < .5$ weights the posterior mass toward smaller models, and visa versa \cite{BMS}. 

\par   The binomial prior discussed above suffers from the fact that the inclusion probability $\theta$ is fixed.  Following \cite{LeySteel2009}, greater flexibility is obtained by treating $\theta$ as random.  A logical choice for the probability distribution of $\theta$ is the Beta distribution with hyperparameters $a, b > 0$, viz.  $\theta \sim \mbox{Beta}(a, b)$, giving rise to the \emph{beta-binomial} model prior 

\par A number of studies have been conducted comparing the performance of combinations of parameter and model priors. It is beyond the scope of this paper to review each of these studies, however, the most recent, and arguably the most comprehensive studies to compare parameter and models priors were conducted by \cite{porwal2022, Porwal2023Effect}.  With respect to parameter priors, \cite{porwal2022}  evaluated 21 widely used methods through extensive simulation studies based on real datasets representing diverse practical scenarios. Three adaptive Bayesian model averaging (BMA) methods emerged as the top performers across all statistical tasks. These methods utilized adaptive versions of Zellner’s $g$-prior, where the prior variance parameter $g$ is either estimated from the data or set as the sample size $n$. Two of the best-performing methods, BMA with $g = \sqrt{n}$ \citep[see][]{fernandezleysteel01a} and local empirical Bayes, were found to be as efficient as LASSO \citep{Tibshirani96}, a commonly preferred variable selection technique. Additionally, BMA outperformed Bayesian model selection, which selects only a single model.

\par With regard to model priors, \cite{Porwal2023Effect} examined eight reference model space priors commonly used in the literature, along with three adaptive parameter priors. To evaluate their effectiveness in variable selection for linear regression models, \cite{Porwal2023Effect} assessed their performance across key statistical criteria, including parameter estimation, interval estimation, inference, and both point and interval prediction. Their analysis was based on an extensive simulation study using 14 real datasets that captured a variety of practical scenarios. \cite{Porwal2023Effect} found that beta-binomial model space priors, defined in terms of the prior probability of model size, performed best on average across different statistical criteria and datasets, surpassing priors that assign equal probability to all models. In contrast, recently introduced complexity priors \citep{Castillo2015} showed relatively weaker performance.

\par For this paper, we examined BMA under a combination of parameter priors (UIP, RIC, BRIC, and HQ) and model space priors (uniform, binomial, and beta-binomial) using the Kullback-Leibler divergence \citep{KullbackLeibler51, Kullback59, Kullback87} as the measure of predictive performance (see Section 6.5). The results showed virtually no difference in terms of Kullback-Leibler divergence across all combinations of parameter and model space priors. Similar results were found in empirical applications by \cite{KaplanHuang2021} and \cite{Kaplan2021Psychometrika}. Therefore, for the purposes of this paper, we utilize the UIP for parameter priors and the uniform distribution for the model space priors.

\subsection{Scoring rules for BMA}

\par A central characteristic of statistics is to develop accurate predictive models \citep{dawid84}.  Indeed, as pointed out by \citep{bernardosmith2000}, all other things being equal, a given model is to be preferred over other competing models if it provides better predictions of what actually occurred.  Thus, a critical element in the development of accurate predictive models is to decide on rules for gauging predictive accuracy -- termed \textit{scoring rules}.  Scoring rules provide a measure of the accuracy of probabilistic projections, and a forecast can be said to be ``well-calibrated'' if the assigned probabilities of the outcome match the actual proportion of times that the outcome occurred.

\par A number of scoring rules are discussed in the literature (see e.g., \cite{Winkler96,bernardosmith2000,JoseNauWinkler08,MerkleSteyvers13,gneitingraftery}), however, for this paper we will primarily evaluate predictive performance under different parameter and model prior settings using the Kullback-Liebler Divergence (KLD) and the \emph{leave-one-out information criterion} (LOOIC). The KLD between two distributions can be written as
\begin{equation}\label{KL}
I(f,g) = \int f(x) log \left(\frac{f(x)}{g(x|\theta)}\right)dx,
\end{equation}
where $I(f,g)$ is the information lost when $g$ is used to approximate $f$. For this paper, the estimated growth rate without predictors is compared to predicted growth rate using BMA along with different choices of model and parameter priors.  The projection model with the lowest KLD measure is deemed best in the sense that the information lost when approximating the ``true'' outcome distribution with the distribution predicted on the basis of the model is lowest. For this paper, LPS values will be obtained using \proglang{BMS}, and the KLD values will be obtained using the \proglang{R} package \proglang{LaplacesDemon} \citep{LaplacesDemon}.

\par \subsection{\textbf{Leave-One-Out Cross-Validation}}
\par For this paper, we utilize \emph{leave-one-out cross validation} (LOO-CV) to evaluate the predictive performance of the three growth curve models discussed earlier. Leave-one-out-cross-validation (LOO-CV) is a special case of $q$-fold cross-validation ($q$-fold CV) when $q=n$.  In $q$-fold CV, a sample is split into $q$ groups (folds) and each fold is taken to be the validation set with the remaining $q-1$ folds serving as the training set.  For LOO-CV, each observation serves as the validation set with the remaining $n-1$ observations serving as the training set. Leave-one-out cross-validation is available in the \proglang{R} software program \proglang{loo} \cite{loo_R}.\footnote{The \textit{widely applicable information criterion} (WAIC) has also been advocated for model selection.  Although the WAIC and LOO-CV are asymptotically equivalent \cite{Watanabe2010}, the implementation of LOO-CV in the \proglang{loo} package is more robust in finite samples with weak priors or influential observations \cite{vehtari2017loo}} 

\par Following \cite{vehtari2017loo} but contextualized in terms of our paper, let $\pi_{1i}$ $(i=1,\ldots,n)$ be an $n$-dimensional vector of slope parameters following a distribution conditional on parameters $\theta$ - viz. $p(\pi_{1}|\theta) = \prod_{i=1}^{n}p(\pi_{1i}|\theta)$. Given a prior distribution on the parameters, $p(\theta)$, we can obtain the posterior distribution, $p(\theta|\pi_{1})$ as well as a posterior predictive distribution of predicted values of growth $\tilde{\pi}_{1}$ written as $p(\tilde{\pi}_{1}|{\pi}_{1}) = \int p(\tilde{\pi}_{1}|\theta)p(\theta|{\pi}_{1})d\theta$. The Bayesian LOO-CV rests on the derivation of the \textit{expected log point-wise predictive density} (ELPD) for new data defined as 
\begin{equation}\label{ELPD}
\mbox{ELPD} = \sum_{i=1}^{n}\int p_t(\tilde{\pi}_{1i})\mbox{log }   
p(\tilde{\pi}_{1i}|{\pi}_{1})d\tilde{\pi}_{1i},
\end{equation}
where $p_t(\tilde{\pi}_{1i})$ represents the distribution of the true but unknown data-generating process for each country's rate of progress toward minimum proficiency $\tilde{\pi}_{1i}$ and where Equation (\ref{ELPD}) is approximated by cross-validation procedures. The ELPD provides a measure of predictive accuracy for the $n$ data points taken one at a time \cite{vehtari2017loo}.  From here, the Bayesian LOO estimate can be written as
\begin{equation}
\mbox{ELPD}_{loo} = \sum_{i=1}^{n}\mbox{log}\,p({\pi}_{1i}|{\pi}_{1-i}),
\end{equation}
where 
\begin{equation}
p({\pi}_{1i}|{\pi}_{1-i}) = \int\,p({\pi}_{1}|\theta)p(\theta|{\pi}_{1-i})d\theta,
\end{equation}
which is the leave-one-out predictive distribution using the log predictive score to assess predictive accuracy. An information criterion based on LOO (LOO-IC) can be derived as
\begin{equation}
\mbox{LOO-IC} = -2\,{\widehat{\mbox{ELPD}}_{loo}}.
\end{equation}
which places the LOO-IC on the ``deviance scale'' (see \citealp{vehtari2017loo} for more details on the implementation of the LOO-IC in \proglang{loo}). Among a set of competing models, the one with the smallest LOO-IC is considered best from an out-of-sample point-wise predictive point of view.

\par As pointed out by \cite{vehtari2017loo}, it can be time-consuming to calculate exact LOO-CV and this may be a reason why LOO-CV is not widely adopted.  To remedy this, \cite{vehtari2017loo} developed a fast and stable approach to obtaining LOO-CV referred to as  \textit{Pareto-smoothed importance sampling} (PSIS-LOO) \citep{vehtari2017loo}. The PSIS approach is implemented in \proglang{loo} \citep{loo_R}.

\section{Data and Workflow}
\subsection{Indicator Selection}

\par To select relevant indicators of countries' rate of progress towards the SDG targets, we extract data from several reliable sources. Following \cite{SDGindex}, we prioritize indicators that are globally relevant, statistically valid, and up-to-date date to attempt to best capture current trends in progress towards the targets in reading and mathematics proficiency across countries. Our choice of indicators originate from the OECD, UNDP, UNESCO, and the World Bank which in turn derive from an idea in systems theory referred to as the \emph{Context-Input-Process-Output} (CIPO) model of national educational systems \citep{Scheerens2023}. A conceptual diagram of the CIPO model is shown in Figure 3.  
\vspace{-.5in}
\begin{figure}[H]
    \centering
    \includegraphics[scale=0.6
]{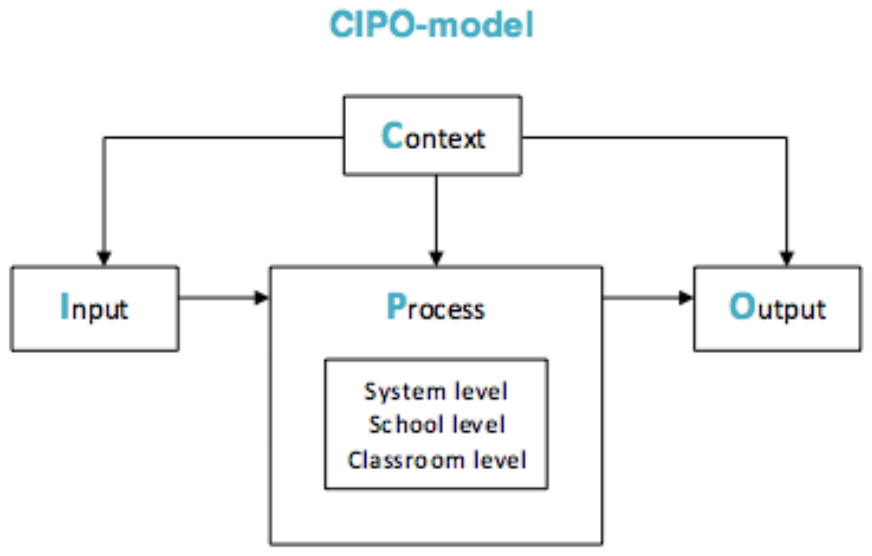}
    \vspace{-3.5in} 
    \caption{Context-Input-Process-Output Model of Scherens (2000)}
    \label{fig:CIPO}
\end{figure}

\par There are many ways to define quality education: Models include quality aspects regarding educational policies (OECD, 2022), school effectiveness (e.g. Scherens, 2000), teacher education, subject specific developments, and individual and social factors (Johnstone et al., 2020). Analysis of quality in education usually takes into account aspects of input, process, and outcomes, depending on the context of the school system (Scheerens et al. 2023) (see Figure \ref{fig:CIPO}). Hence, quality of education cannot be defined by outcomes alone, but must take into account contextual factors, input resources and processes which form the education system on different levels, leading to education outcomes and also can be seen as prerequisites for equity in education. 
\par When analyzing the concrete SDG targets in relation to the CIPO model (see Table \ref{tab:stackmodels}), it becomes clear that almost all of them either refer to outcome of education in the sense of the goals that are to be reached, or to aspects of access of education which can be seen as input factors or resources provided by an education system. Only target 4.c referring to a qualified teacher workforce could both be seen as an Input indicator but might also be related to the educational processes in schools under the assumption that qualification is needed to ensure quality in the education setting. However, it can also be seen as in input factor as teacher qualification requires resources invested into teacher education itself.

Indicators relating to the context of education systems are not specifically stated under the SDG 4 targets. It can be assumed that context factors are rather country specific (like the structure of the school system) also might be changing rapidly (like flows of immigration or specifications for student admission). Moreover, indicators of context factors can also be found among the targets of other SDGs, for example indicators of poverty (SDG Goal 1). 

Following the aspects of the CIPO model and linking it to the idea of predicting change in educational outcomes, respective indicators for input, context and processes of the education system would be needed for all countries reporting on their progress on the SDG. Hence, also when thinking about predicting progress towards the goals, indicators of input, context and processes need to be considered. 

In addition to using the very few indicators already included in goal 4 of the SDG, the CIPO model can be used as a foundation to select additional indicators from other data sources which indicate aspects relevant to describe quality in education. These indicators of context, input, and processes can then be used to predict the change of the rate of progress towards the outcome indicators. Constructs being assessed in ILSA can provide the respective data.

\par After choosing our relevant indicators, we created our analytic dataset using the difference between 2022 and 2009 values for each indicator. We used difference variables since we are estimating the rate of progress in percent minimum proficiency, where we are interested in the change over time, not just the starting minimum proficiency percentage. Changes in policy, social conditions, and other relevant conditions can lead to changes in the rate of growth in reading and mathematics proficiency \citep{Kyriakides2017, Strietholt2019}. At the time of this writing many countries did not have data for 2022, so we substituted 2021 data when necessary. A complete description of the indicators chosen, what years we used to form the difference variables, and the breakdown of Bayesian stacking models is contained in Table \ref{tab:stackmodels}. 

\begin{table*}[t]
\centering
\caption{List of indicators used for BMA categorized in terms of elements of the CIPO model.}
\label{tab:stackmodels}
\resizebox{\textwidth}{!}{  
\begin{tabular}{l|c|c}
\toprule
\textbf{Indicator} & \textbf{Member Models for Stacking} & \textbf{Difference Years} \\ 
\midrule
Employment females as percentage of employment annual & Context (Gender Gap) & 2021 - 2009 \\
Gender development index & Context (Gender Gap) & 2022 - 2009 \\
Gender gap index GEQ & Context (Gender Gap) & 2022 - 2009 \\
Labor force participation rate female percentage of female population & Context (Gender Gap) & 2021 - 2009 \\ 
\midrule
Adolescents out of school of lower-secondary age & Outcome/Input (Education) & 2022 - 2009 \\
Children out of school of primary school age & Outcome/Input (Education) & 2022 - 2009 \\
Gross enrollment ratio primary both sexes & Outcome/Input (Education) & 2022 - 2009 \\
Gross enrollment ratio secondary both sexes & Outcome/Input (Education) & 2022 - 2009 \\
Lower-secondary school starting age years & Outcome/Input (Education) & 2022 - 2009 \\
Official entrance age to lower-secondary education years\textsuperscript{1} & Outcome/Input (Education) & 2022 - 2009 \\
Official entrance age to pre-primary education years & Outcome/Input (Education) & 2022 - 2009 \\
Official entrance age to primary education years & Outcome/Input (Education) & 2022 - 2009 \\ 
\midrule
GDP (standardized) & Context (SES) & 2021 - 2009 \\
Human development index & Context (SES) & 2022 - 2009 \\
Index highest occupational status of parents & Context (SES) & 2022 - 2009 \\
Rural population as percentage of total population & Context (SES) & 2022 - 2009 \\ 
\midrule
Expected years of schooling & Context (Education) & 2022 - 2009 \\
Funding government & Context (Education) & 2022 - 2009 \\
Government expenditure on education percentage of GDP & Context (Education) & 2022 - 2009 \\
Government expenditure on education percentage of government expenditure & Context (Education) & 2022 - 2009 \\
Index school size & Context (Education) & 2022 - 2009 \\ 
\midrule
Percentage of full-time teachers per school & Processes/Input (Education) & 2022 - 2009 \\
Percentage of part-time teachers per school & Processes/Input (Education) & 2022 - 2009 \\
Number of class periods in mathematics & Processes/Input (Education) & 2022 - 2009 \\
Primary school starting age years\textsuperscript{1} & Processes/Input (Education) & 2022 - 2009 \\
Teaching hours lower secondary & Processes/Input (Education) & 2021 - 2009 \\
Teaching hours primary & Processes/Input (Education) & 2021 - 2009 \\
Teaching hours upper-secondary & Processes/Input (Education) & 2021 - 2009 \\ 
\midrule
Percentage of teachers in pre-primary education who are female & Context (Education) & 2022 - 2009 \\
Percentage of teachers in primary education who are female & Context (Education) & 2022 - 2009 \\
Percentage of teachers in secondary education who are female & Context (Education) & 2022 - 2009 \\ 
\bottomrule
\end{tabular}}
\vspace{2pt}
\begin{minipage}{\textwidth}
\scriptsize
\raggedright
\textsuperscript{1} Denotes indicators removed from model averaging due to colinearity.
\end{minipage}
\end{table*}

\par The first step of the workflow for this paper begins with estimating the rate of progress in the percentage of students meeting or exceeding minimum proficiency in reading and mathematics for both boys and girls separately. Our approach to the estimation of these growth curve models recognizes that minimum proficiency is a proportion that varies continuously across countries and therefore is not normally distributed as linear growth curve modeling assumes.  Therefore, we address this problem by treating minimum proficiency as having been generated from a $beta$($a$, $b$) distribution, where $a$ and $b$ are shape and scale parameters, respectively \citep[see e.g.][]{kaplanbayesbook2}.  The $beta$ distribution is appropriate for data in the range [0,1]. 

\par In line with \cite{KaplanHarraStampkaJude2025}, three models are estimated on the logit transformed data: (a) a linear growth curve model, (b) a latent basis model allowing the last latent basis term (2022) to be estimated by the data in order to examine the potential impact of the pandemic on the proportion of students in countries achieving minimum proficiency, and (c) a latent basis model that allows the last two latent basis terms (2018 and 2022) to be free in order to capture changes in the percentage achieving minimum proficiency that were already occurring prior to the pandemic. We evaluate each model using the LOOIC and KLD and choose one of these models for the next step in the workflow.

\par In the second step of the workflow, we adopt an $\mathcal{M}$-closed framework and utilize BMA to predict the percentage of students achieving or exceeding minimum proficiency. Again, because we need to account for the non-normality of minimum proficiency we apply BMA to the logit of the minimum proficiency values.  For this analysis, we also examine the sensitivity of the results to changes in the initial starting conditions of the analysis by examining alternative model and parameter priors as discussed in Section 6.3 and Appendix A.  


\par In the third step of the workflow, we will use the model averaged coefficients from the BMA prediction of the logit of rate-of-change to develop out-of-sample predictions of the percentage of students achieving minimum proficiency by plugging in the optimally predicted growth rate into our growth model and projecting the next two cycles of PISA, namely 2029 and 2033.  At this point, we transform the predicted results back to 0-100 scale in order to obtain plots.  Plots will be provided for the average across all countries and for each country separately. Four case study countries will be provided.

\subsection{Missing Data Procedures}

\par Given the numerous data sources, years of interest, and countries included in our analysis, we encountered some missing data due to some countries not collecting or reporting data for our chosen indicators. To maintain as large of a sample with as many relevant indicators as possible, our missing data imputation procedure is as follows.

\par Due to initial difficulty achieving convergence with our multiple imputation algorithm, when available, we substituted a neighboring year's observed data instead of solely relying on predictive mean matching for missing data. For example, the indicator "gross enrollment ratio primary school (both sexes)" only has reported data for eight of our selected 53 countries in 2022 at the time of this writing. Meanwhile, 50 of 53 countries reported data in 2021. The observed 2021 values were substituted for 2022 when available, and then we used multiple imputations to impute the rest. After the substitutions, only three countries required imputation for this indicator as opposed to 50 originally missing. This procedure ensured we could maximize the number of indicators included in our analysis. Despite participating in all 2009-2022 PISA cycles, Chinese Taipei was removed from our analytic dataset due to missing data on almost all the included indicators, which made multiple imputation algorithm convergence challenging. Once all available observed substitute data were used, we then imputed the remaining missing values using \emph{predictive mean matching} \citep{rubin86fileconcat} via the \proglang{R} package \proglang{mice} \citep{mice}. 

\par Predictive mean matching produces unbiased estimates under the assumption that the data are missing completely at random (MCAR) or missing at random (MAR) \citep{littlerubin2019}. In the present analysis, we assume the data are missing at random.  Predictive mean matching imputes missing values by matching the predicted values from the observed data using a predictive mean metric to the predicted values with regression imputation. Next, the observed value is used for the imputation. So for each regression model, there is a predicted value for both observed and missing data. The predicted value for the observed data is then matched to the predicted missing data value, for example using a nearest neighbor metric. After a match is made, the missing value is replaced by the observed value (as opposed to the predicted value). Random matches are chosen when multiple matches are found. 

\par To best account for uncertainty surrounding the imputation, multiple draws for missing values should be made. However, we only conducted this process for a single imputed data point to ensure model convergence. While using multiple imputations is considered best practice for producing the most reliable results, we argue that using a single imputation via predictive mean matching is a significant improvement of listwise deletion or using a much smaller set of complete indicators.

\section{Results}
\par This section presents the results in alignment with the workflow described in Section 7.
\subsection{Growth modeling results}
\par Table \ref{tab:GCM} presents growth curve modeling results under three specifications: (a) the linear model (M0), (b) latent basis model  (M1) wherein the last time point is freely estimated, allowing a data-based estimation of the potential impact of COVID, and (c) latent basis model 2 (M2) which allows the last two time points to be estimated under the assumption that the decline in the percentage reaching minimum levels of proficiency occurred before the onset of the pandemic.

\begin{table*}[hbt!]
\centering
\caption{Posterior estimate of starting percent minimum proficiency and rates of progress, 95\% credible intervals (in parentheses), and predictive evaluation under linear and two latent basis models.}
\label{tab:GCM}
\resizebox{\textwidth}{!}{ 
\begin{tabular}{lcc|cc}
\toprule
 & \multicolumn{2}{c|}{\textbf{Reading}} & \multicolumn{2}{c}{\textbf{Mathematics}} \\ 
\cmidrule(lr){2-3} \cmidrule(lr){4-5}
 & \textit{Boys} & \textit{Girls} & \textit{Boys} & \textit{Girls}  \\ 
\midrule
\textbf{Starting minimum proficiency percentage} &  &  &  &    \\
\hspace{0.2cm} Linear model M0 & 68.24 (65.05, 71.51) & 82.45 (80.08, 84.67) & 70.97 (67.57, 74.07) & 69.93 (66.27, 73.16) \\
\hspace{0.2cm} Latent basis M1 & 68.22 (64.97, 71.41) & 82.44 (79.90, 84.54) & 70.99 (67.70, 74.30) & 69.86 (66.36, 73.25) \\
\hspace{0.2cm} Latent basis M2 & 68.23 (65.05, 71.38) & 82.44 (80.00, 84.54) & 71.09 (67.80, 74.25) & 69.98 (66.60, 73.48) \\
\midrule
\raggedright\textbf{Rate of progress} &  &  &  &    \\
\hspace{0.2cm} Linear model M0 & -1.10 (-2.11, -0.11) & -4.07 (-5.11, -3.02) & -1.30 (-2.37, -0.19) & -0.17 (-1.63, 1.30) \\
\hspace{0.2cm} Latent basis M1 & -1.10 (-2.08, -0.06) & -4.00 (-4.97, -3.01) & -1.40 (-2.51, -0.31) & 0.16 (-1.69, 1.84) \\
\hspace{0.2cm} Latent basis M2 & -1.09 (-2.05, -0.05) & -4.02 (-5.14, -2.96) & -1.87 (-3.03, -0.62) & -0.31 (-2.45, 1.71) \\
\midrule
\raggedright\textbf{Predictive evaluation} &  &  &  &    \\
\hspace{0.2cm} Linear model M0 & 215.57 & 259.52 & 207.30 & 237.58 \\
\hspace{0.2cm} Latent basis M1 & 215.68 & 258.55 & 207.08 & 237.43 \\
\hspace{0.2cm} Latent basis M2 & 215.77 & 259.44 & 203.99 & 239.32 \\ \hline
\bottomrule
\end{tabular}
} 
\scriptsize
\vspace{2pt}
\begin{minipage}{\textwidth}
\scriptsize
\raggedright

\vspace{4pt}

Predictive performance is compared using LOO-IC estimates for each condition.
\end{minipage}

\end{table*}

\par For boys and girls both we observed virtually no difference between the models according to the LOO-IC predictive criterion. An inspection of the estimated starting percent minimum proficiency and rate of progress also showed no substantive differences, with the possible exception of the rate of progress in mathematics minimum proficiency for girls under latent basis M1 and M2 relative to the linear model. For the remainder of the paper, we will use estimates of the rate of progress in meeting minimum proficiency under latent basis M1, again to allow for any possible impact of the pandemic.

\subsection{Bayesian model averaging results}
\par Bayesian model averaging was applied to the 31 predictors shown in Table \ref{tab:stackmodels} using the \proglang{R} program \proglang{BMS} \citep{BMS}. The analysis utilized unit information priors for all model parameters and uniform model priors. 
To begin, Table \ref{tab:topmodels} presents the posterior model probabilities (PMPs) for the top 5 models explored by the BMA algorithm as well as the total PMP across all models explored in the model space.
\begin{table}[hbt!]
\centering 
\caption{Posterior model probabilities for the top five models along with total posterior model probability across all models.}
\begin{tabular}{lcc|cc}

\toprule
 & \multicolumn{2}{c|}{\textbf{Reading}} & \multicolumn{2}{c}{\textbf{Mathematics}} \\ \hline
 & \textit{Boys} & \textit{Girls} & \textit{Boys} & \textit{Girls} \\ \hline

Model 1 & 0.04 & 0.03 & 0.03 & 0.04 \\
Model 2 & 0.02 & 0.02 & 0.02 & 0.02 \\
Model 3 & 0.01 & 0.02 & 0.02 & 0.02 \\
Model 4 & 0.01 & 0.02 & 0.02 & 0.02 \\
Model 5 & 0.01 & 0.02 & 0.02 & 0.02 \\ \hline
\textit{Total PMP} & \textit{0.54} & \textit{0.47} & \textit{0.56} & \textit{0.54} \\ \bottomrule

\end{tabular}
\label{tab:topmodels}
\end{table}

\noindent We find that the total PMPs for both reading and mathematics and across boys and girls are well below 1.0 suggesting substantial model uncertainty and thus justifying the use of BMA over the selection of one specific model.  Indeed the top model (Model 1) which would be the model selected on the basis of the Bayesian information criteria illustrates the extent of model uncertainty.

\par Tables \ref{tab:BMA_reading} and \ref{tab:BMA_math} present the results for the top 6 predictors of the rate of progress in the percentage of boys and girls at or above the PISA minimum proficiency for reading and mathematics, respectively. 
The column labeled PIP shows the posterior inclusion probabilities, which represents the proportion of the total posterior model probability (or model mass) that includes that predictor across all models considered in the averaging process. For example, we find that for boys, the top predictor of the rate of progress in reading minimum proficiency is the change from 2009 to 2022 (or 2021) in the percentage of teachers in pre-primary education who are female.  However, the PIP is 0.78, meaning that 78\% of the total posterior model probability mass includes that predictor. It is arguble whether this predictor is important in a substantive sense.  Contrast this with the rate of progress in minimum proficiency for girls in mathematics.  Here we find that 96\% of the total probability mass contains the predictor measuring the change from 2009 to 2022 (2021) in the percent of children out of school of primary school age.  Arguably, this predictor is important for assessing the rate of progress in mathematics minimum proficiency for girls. 
This latter finding notwithstanding, across domains and for boys and girls we find that few of the predictors used in this study are clearly important in predicting the logit of the rate of progress in reading and mathematics. Although we believe that we have captured a considerable number of the most important predictors of the rate of progress in minimum proficiency driven by our understanding of the CIPO model, there is considerable room for future theory and indicator development that might improve PIP levels.

\begin{table}[!ht]
\centering
\caption{BMA results for reading proficiency. Note that the outcome of interest is the estimated rate of progress (i.e. growth rate) in the percentage of students at or above the PISA minimum proficiency. All indicators are measured as the difference between 2022 (or 2021) and 2009 values.}
\label{tab:BMA_reading}
\begin{tabular}{lc} 
\toprule
\multicolumn{2}{c}{\textbf{Boys}} \\ 
\midrule
 & PIP \\
Percentage of teachers in pre-primary education who are female & 0.78 \\
Rural population percent of total population & 0.52  \\
Percentage of full-time teachers per school & 0.43 \\
Teaching hours upper-secondary & 0.36 \\
Teaching hours primary & 0.31 \\
Employment females as percent of employment annual &  0.30 \\ 
\midrule
 &  \\
\multicolumn{2}{c}{\textbf{Girls}} \\ 
\midrule
 & PIP \\
Employment females as percentage of employment annual & 0.56 \\
Gender development index & 0.51  \\
Percentage of teachers in pre-primary education who are female & 0.50 \\
Children out of school of primary age & 0.41 \\
Teaching hours upper-secondary & 0.40 \\
Official entrance age to primary education years & 0.38 \\
\bottomrule
\end{tabular}
\scriptsize
\vspace{1pt}
PIP: Posterior inclusion probability.
\end{table}

\begin{table}[!ht]
\centering
\caption{BMA results for mathematics proficiency. Note that the outcome of interest is the estimated rate of progress (i.e. growth rate) in the number of students at or above the PISA minimum proficiency. All indicators are measured as the difference between 2022 (or 2021) and 2009 values.}
\label{tab:BMA_math}
\begin{tabular}{lc} 
\toprule
\multicolumn{2}{c}{\textbf{Boys}} \\ 
\midrule
 & PIP \\
Children out of school of primary-school age & 0.90 \\
Index highest occupational status of parents & 0.67 \\
Percentage of full-time teachers per school & 0.66  \\
Employment females as percentage of employment annual & 0.46 \\
Percentage of pre-primary teachers who are female & 0.45 \\
Teaching hours lower secondary & 0.41  \\ 
\midrule
 &  \\
\multicolumn{2}{c}{\textbf{Girls}} \\ 
\midrule
 & PIP \\
Children out of school of primary-school-age & 0.96  \\
Employment females as percentage of employment annual & 0.48 \\
Teaching hours upper-secondary & 0.42 \\
Index school size & 0.42  \\
Percentage of part-time teachers per school & 0.40 \\
Index highest occupational status of parents & 0.38 \\
\bottomrule
\end{tabular}
\scriptsize
\vspace{1pt}
PIP: Posterior inclusion probability
\end{table}

\section{Overall country projections and case studies}
We begin by presenting BMA-based projections for reading and mathematics and for boys and girls separately across all countries/economies used in this analysis.

\subsection{Overall country projections}
\par
\begin{figure}[h]
    \centering
    \includegraphics[width=.9999\linewidth]{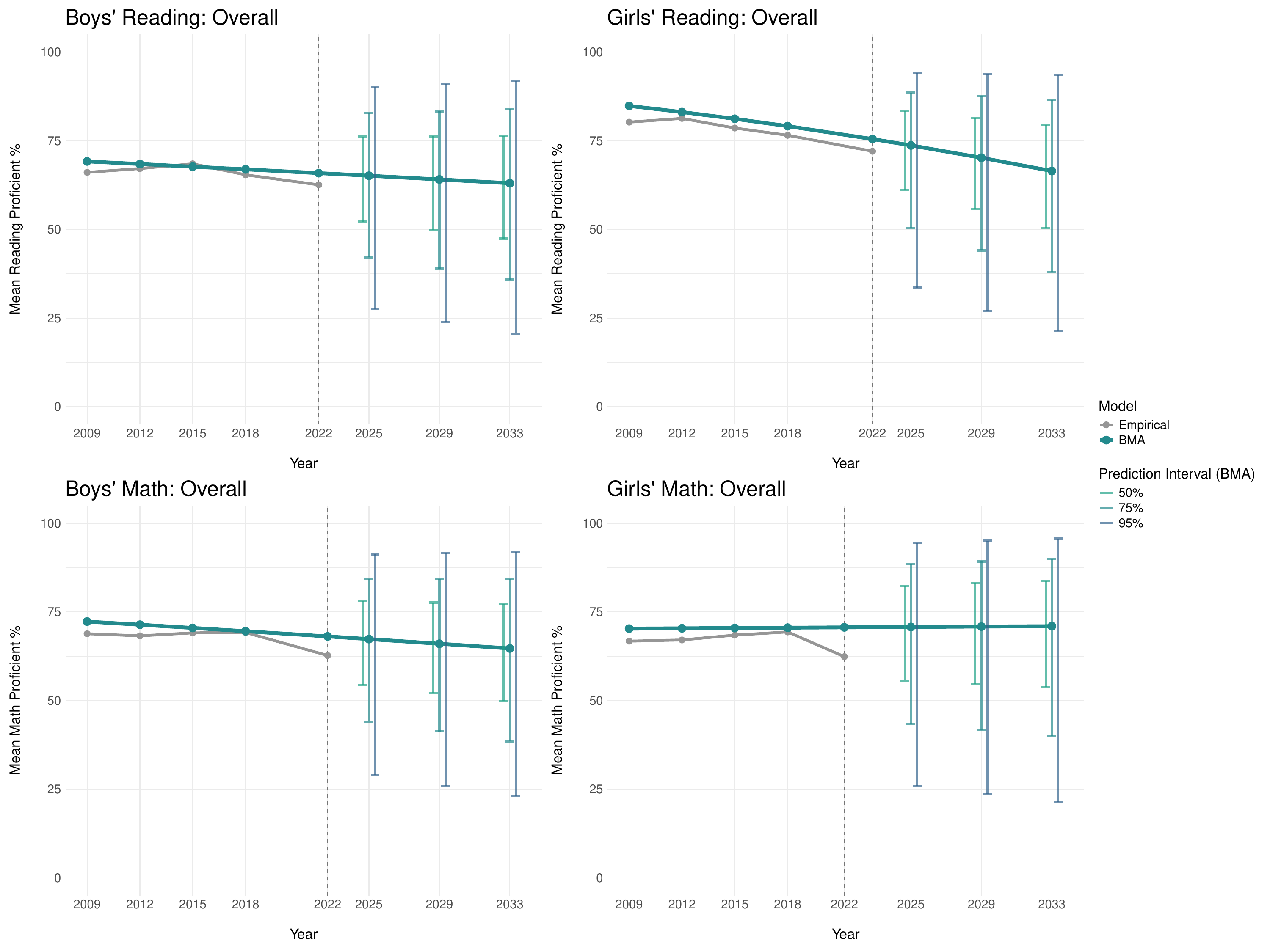}
    \caption{BMA-based projections and prediction intervals for mathematics (a) and reading (b) for boys and girls. Note that now PISA is on an every four-year cycle.}
    \label{fig:trajectories}
\end{figure}

 Overall country projections are shown in Figure \ref{fig:trajectories}.

We find that across the countries used in this analysis, there has been relatively small amounts of change in the percentage of students who have reached or exceeded the PISA-defined minimum proficiency level and that, moreover, the projection to 2033 appears to be relatively stable at about 65\% at or above minimum proficiency across these countries. Another way to consider this finding is that without intervention, an average of approximately 35\% of 15-year-old students across these countries will remain below minimum proficiency in reading and mathematics beyond 2030, independent of gender.

\subsection{Case studies}

In this section, we present the results of four case studies: Croatia, Germany, the United Kingdom, and the United States.  These countries were chosen to illustrate the differences in the predictive quality of the model across boys and girls and across reading and mathematics. Remaining countries are provided in the supplementary material.

First, to assess the quality of the prediction models for our case studies, we utilize  \emph{BMA predictive densities} as implemented in \proglang{BMS} \citep{BMS}. These predictive densities provide a means of checking the prediction model of the growth rate based on BMA against the unconditional growth rate based on latent basis M1, allowing examination of problems with the prediction model and/or potential outlier countries.\footnote{This analysis assumes that latent basis model M1 is very close to the actual growth rate.  This might be a reasonable assumption given the data based fitting of M1 and the fact that this model did not differ compared to the linear model and latent basis model M2.} In the upper panels of the figures, the dashed vertical line is the unconditional growth rate and the solid line is the model-predicted (expected) growth rate based on the predictor information from the country of interest and the information from all other countries.  Thus, our approach borrows information from all the countries in providing a prediction of the growth rate for the country of interest. The 95\% quantiles for the predictive densities are also displayed. 

\par Second, in the lower panel of the following figures, we present the trajectory plots of minimum proficiency for the case study countries, and for boys and girls and proficiency domains.  The gray line is the empirical percentage of the countries students at or above the PISA level 2 minimum proficiency and the green line is the estimated trajectory under latent basis model M2, with the estimated intercept and growth rate obtained from BMA. We also describe change in the percentage of students at or above minimum proficiency from 2009 to the projected value in 2033 and these are show in Table \ref{tab:projections} below.

\begin{table*}[htbp]
\centering
\caption{Projections of the percent of students meeting minimum proficiency for the case study countries by gender and literacy domain.}
\begin{tabular}{lcccccc}

\toprule
               & \multicolumn{6}{c}{\textbf{Reading}}                                   \\ 
\cmidrule(lr){2-7}
               & \multicolumn{3}{c}{\textit{Boys}} & \multicolumn{3}{c}{\textit{Girls}} \\
\midrule
               & 2009      & 2033      & \% Change    & 2009      & 2033      &  \% Change     \\
Croatia        & 68     & 59    &  -9    & 88     & 66     & -22     \\
Germany        & 76    & 70     &  -6    &  87    & 67     &  -20     \\
United Kingdom & 77    & 76     &  -1    &  86    & 80     & -6       \\
United States  & 79    & 66     &  -13    &  86    & 68     &  -18   \\
\midrule
               & \multicolumn{6}{c}{\textbf{Mathematics}} \\ 
\cmidrule(lr){2-7}
               & \multicolumn{3}{c}{\textit{Boys}} & \multicolumn{3}{c}{\textit{Girls}} \\ 
\midrule
               & 2009      & 2033      &  \% Change    & 2009      & 2033      &  \% Change     \\
Croatia        &  69   & 62    &  -7    & 65     & 64     &  -1    \\
Germany        & 83    & 71    &  -12    & 80     & 73     &  -7     \\
United Kingdom & 82    & 75    &  -7    & 77     & 82     &    +5   \\
United States  & 80    & 59    &  -21    & 73     & 67     &   -6   \\
\bottomrule

\end{tabular}%
\scriptsize
\vspace{1pt}
\raggedright 

Note that 2033 is a projection based on the latent basis model M1.
\label{tab:projections}
\end{table*}


\subsection{Case study 1: Croatia} 
\begin{figure}[H]
\centering
  {\includegraphics[scale=.23]{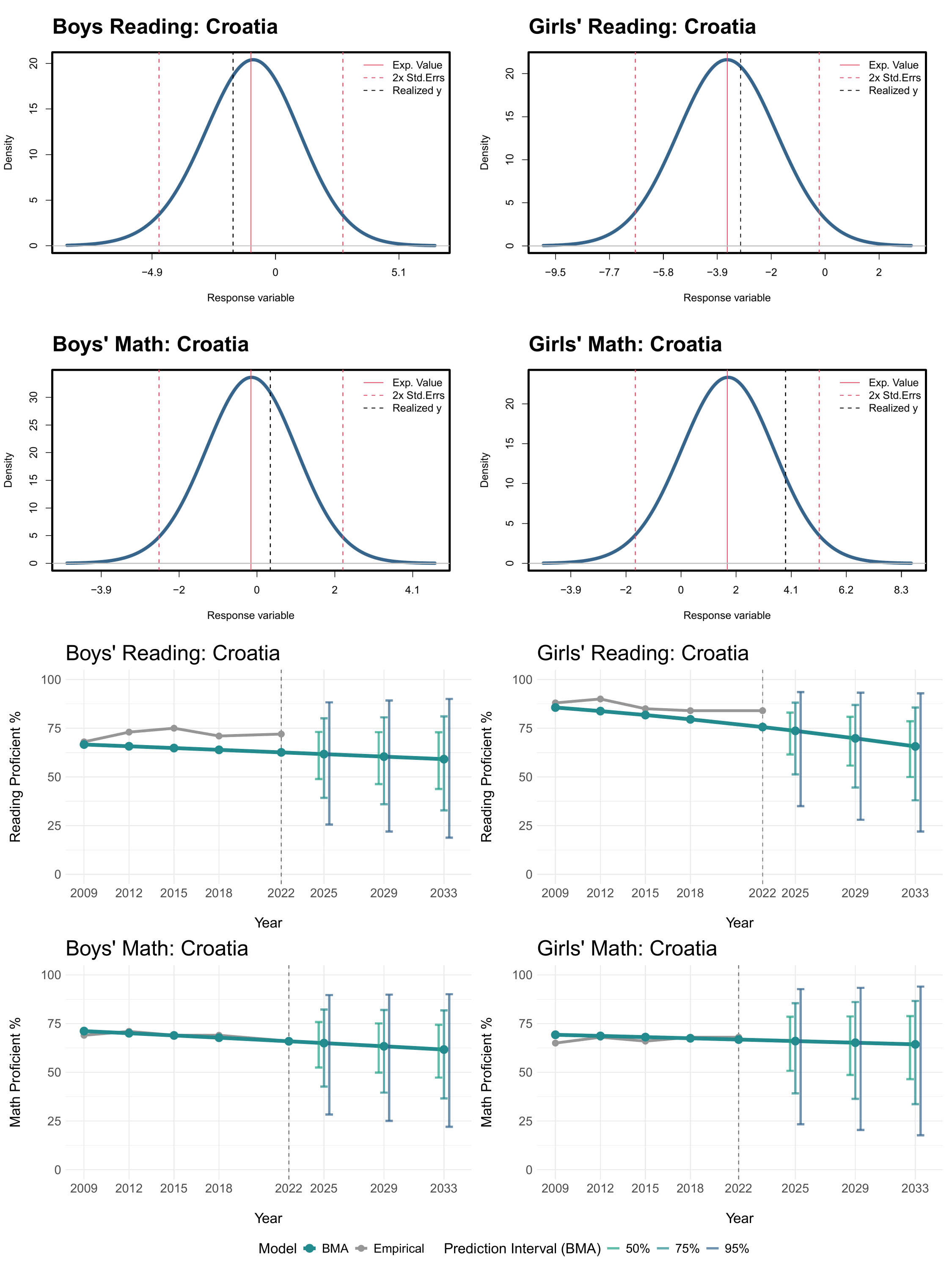}
  \caption{Predictive density plots (above) and trend lines (below) for percentage of Croatian students at the minimum proficiency level in mathematics from 2009 to 2022.}\label{fig:Croatia}}
\end{figure}

Predictive density plots and trend plots for Croatia are displayed in Figure \ref{fig:Croatia}. For boys and girls and for both proficiency domains, the BMA predictive growth rates are all within the 95\% quantiles of the unconditional growth rate, however, we find that the fit is not excellent, particularly for boys in reading and for girls in reading and mathematics. 

\par The trajectory plots reveal that the trend is mostly constant for all cases except for a slightly more noticeable decline for girls' in the minimum proficiency percentage in reading over time. Inspecting Table \ref{tab:projections}, in 2009, the minimum proficiency percentage for Croatian boys in reading was 68\% and in 2033 it is projected to be 58\%, a drop of approximately 9\%. For Croatian girls in reading, the projected decline is much greater with 88\% of the girls at or above minimum proficiency in 2009 but projected to be at 66\% by 2033 for a drop of 22\%.  Thus while the percent of Croatian girls at minimum proficiency is uniformly greater than that for boys, the decline is much greater for girls.   
\par The minimum proficiency percentage for Croatian boys in mathematics is projected to drop by 7\% in 2033 compared to where it was in 2009. For Croatian girls in mathematics, the projected decrease in the minimum proficiency percentage is much smaller at 1\%.

\subsection{Case study 2: Germany}
\par
\begin{figure}[H]
\centering
  {\includegraphics[scale=.23]{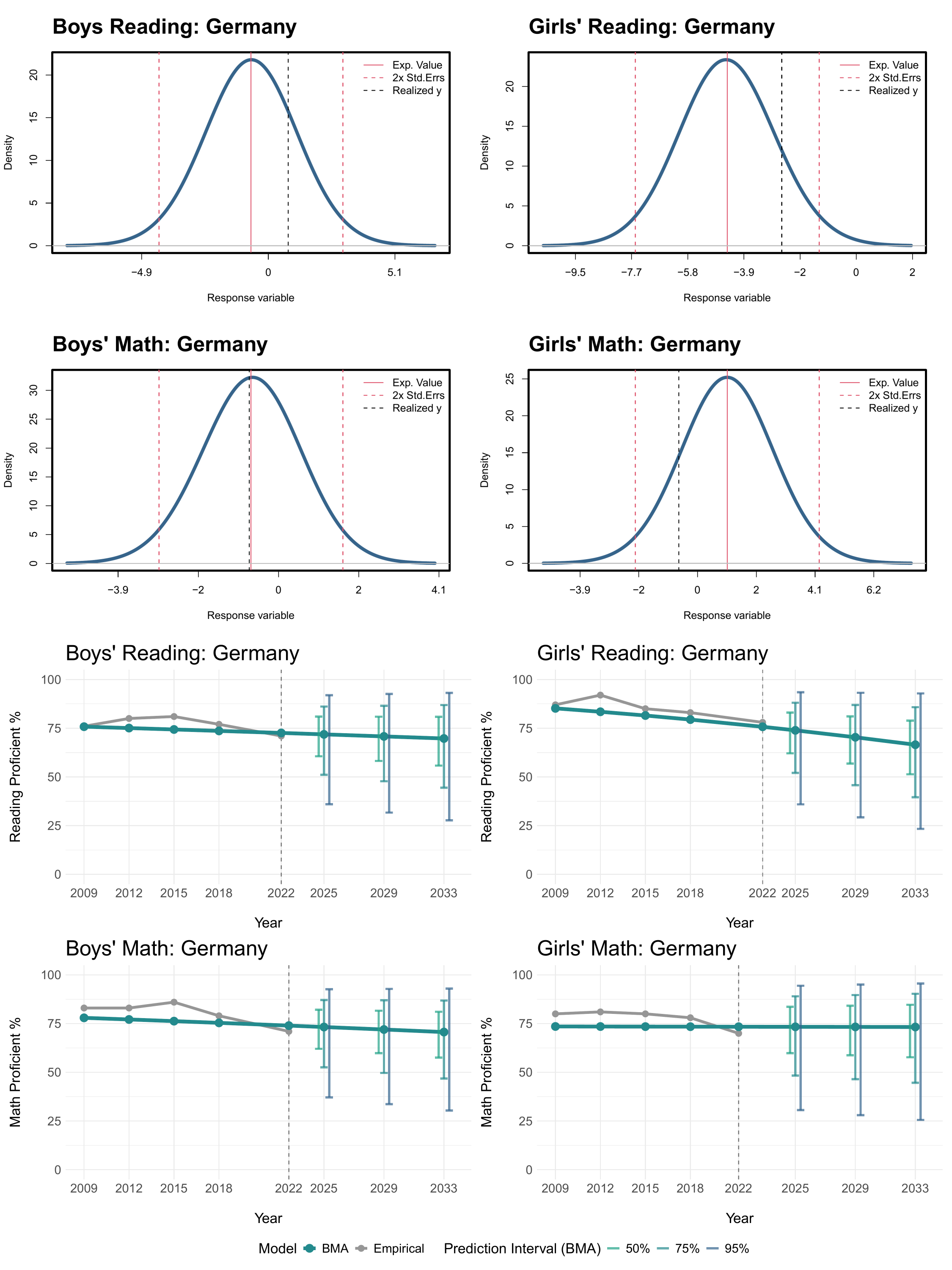} 
  \caption{Predictive density plots (above) and trend lines (below) for percentage of German students at the minimum proficiency level in mathematics from 2009 to 2022.}\label{fig:Germany}}
\end{figure}

 Predictive density plots and trend plots for Germany are displayed in Figure \ref{fig:Germany}. For math, model fit was quite good for boys and girls, whereas for reading the model fit was not as good.  Nevertheless, in all four cases, the BMA predicted rate of progress was within the 95\% prediction interval of the actual growth rate. 

\par Inspection of the trajectory plots for Germany as well as Table \ref{tab:projections} reveal a more noticeable decline in the percent of German girls meeting or exceeding minimum proficiency in reading compared to German boys. In 2009, the percent of German boys reaching or exceeding minimum proficiency in reading was 76\% and projected to be 70\% in 2033 for a drop of 6\%. In contrast, although the percent of German girls reaching or exceeding minimum proficiency is greater than German boys in 2009, the decline is much greater at 20\%.

\subsection{Case study 3: United Kingdom}
\par Predictive density plots and trend plots for the United Kingdom are displayed in Figure \ref{fig:UK}. 
From Table \ref{tab:projections}, the percentage of UK boys reaching or exceeding minimum proficiency for reading in 2033 is predicted to be 76\%, a 1\% drop compared to 2009.  For UK girls, the projected percentage meeting or exceeding minimum proficiency in 2033 is 80\% but representing a larger drop of 6\% from 2009 compared to boys.  

\par Regarding mathematics, the percent of UK boys meeting or exceeding minimum proficiency in 2033 is 75\% representing a 7\% drop from the 2009 baseline percentage.  For UK girls, we predict a 5\% increase.

\begin{figure}[H]
\centering
  {\includegraphics[scale=.23]{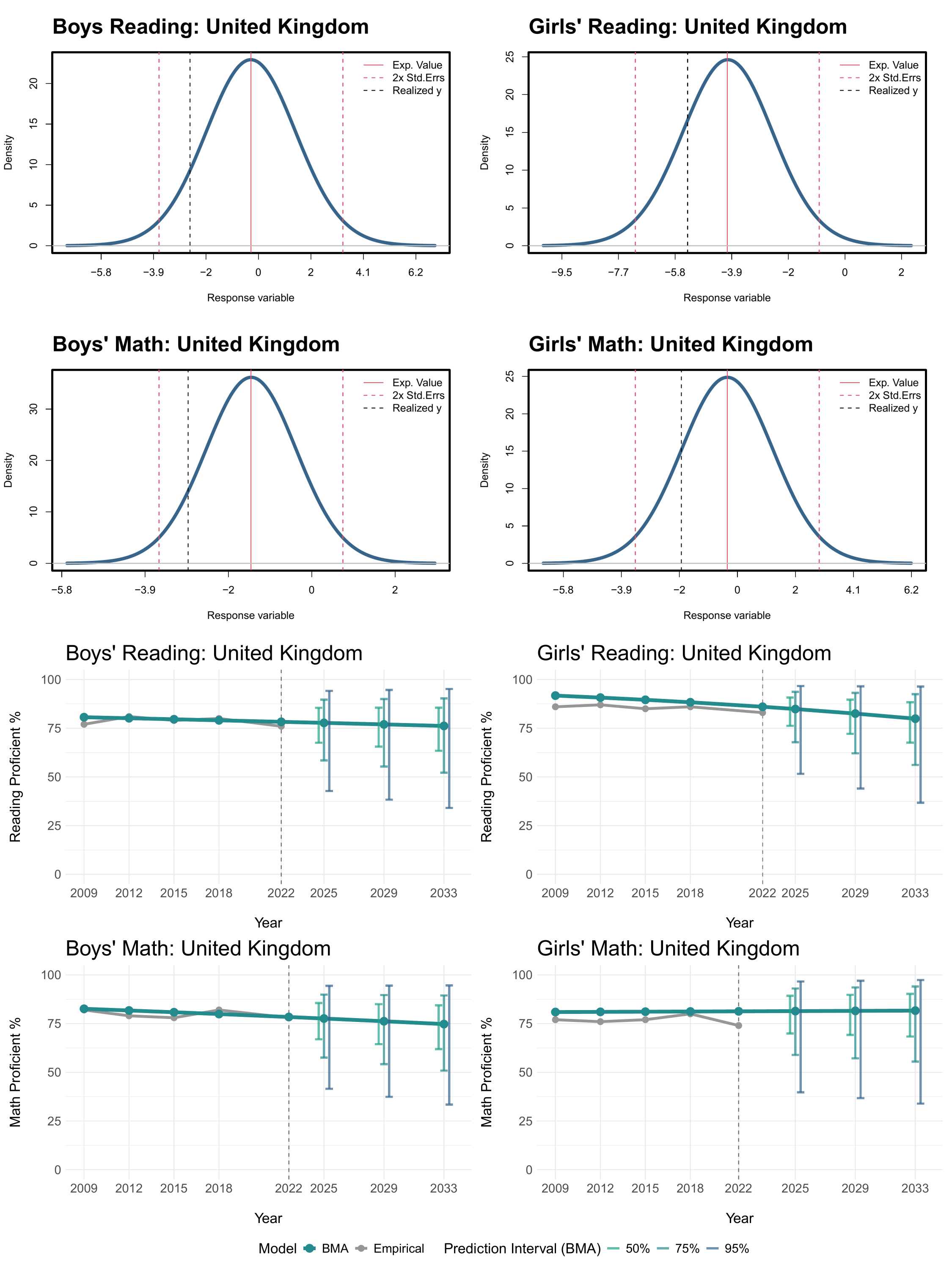} 
  \caption{Predictive density plots (above) and trend lines (below) for percentage of UK students at the minimum proficiency level in mathematics from 2009 to 2022.}\label{fig:UK}}
\end{figure}

\subsection{Case Study 4: United States}
\par Finally, in Table \ref{tab:projections} for the United States, the projected percentage of boys meeting or exceeding minimum proficiency in reading in 2033 is 66\% representing a 13\% drop compared to the 2009 baseline of 79\%.  For US girls, the projected percentage meeting or exceeding minimum proficiency is 68\% representing an 18\% decline relative to the baseline 2009 percentage. This decline is despite the fact that girls start off at an advantage over boys in reading and maintain that advantage in 2033. 

\par Regarding mathematics, we find that the percent of US boys projected to meet or exceed minimum proficiency in 2033 is 59\% representing an 21\% decline from the 2009 baseline.  We find that the percent of US girls projected to meet or exceed minimum proficiency in 2033 67\% representing a lesser decline of 6\% from the 2009 baseline.

\begin{figure}[H]
\centering
  {\includegraphics[scale=.23]{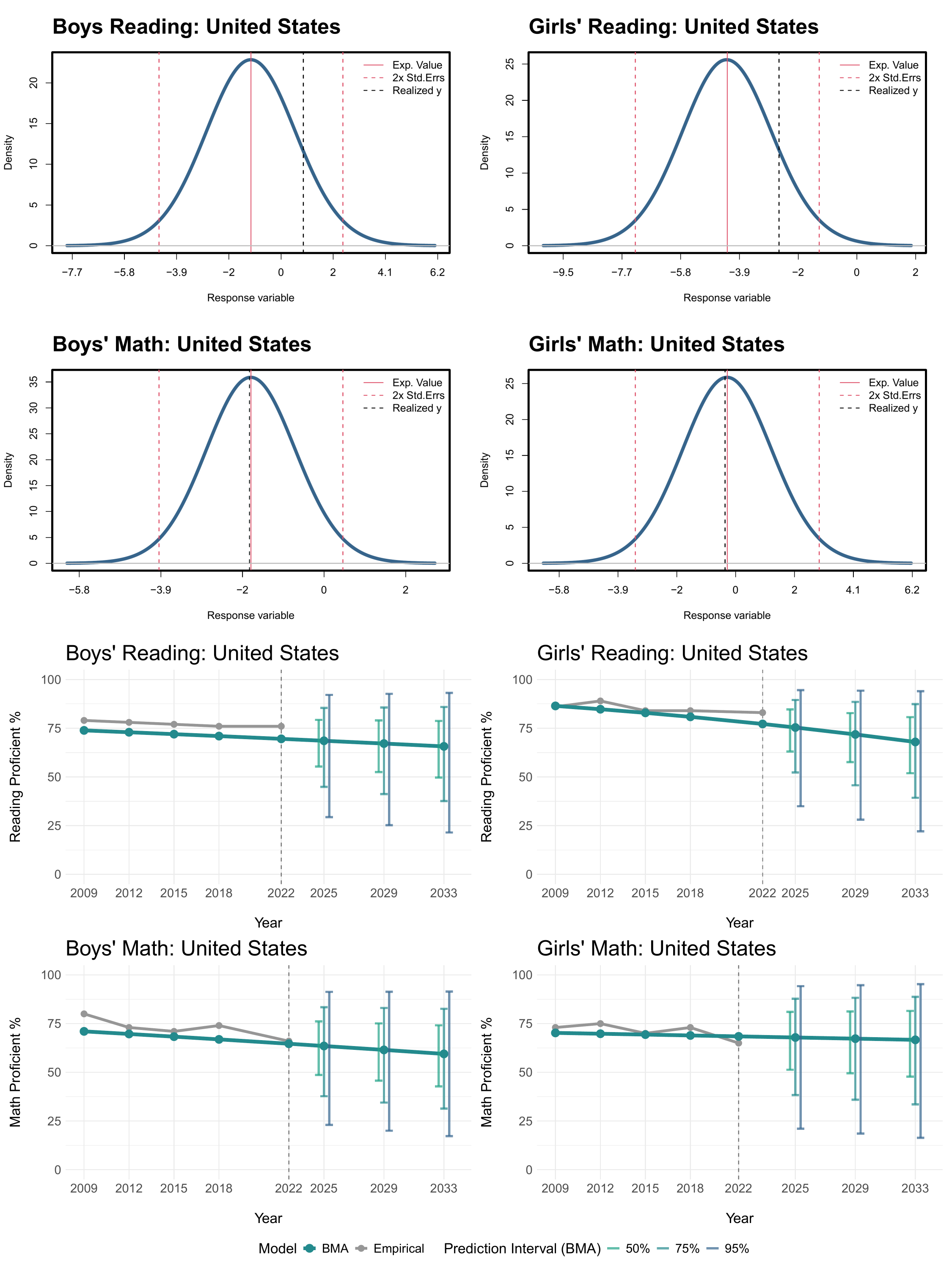} 
  \caption{Predictive density plots (above) and trend lines (below) for percentage of US students at the minimum proficiency level in mathematics from 2009 to 2022.}\label{fig:UK}}
\end{figure}

\section{Conclusions}
\par Using country-level trend data from PISA, this paper provided probabilistic projections of the percent of students in a country meeting PISA-defined minimum proficiency levels in reading and mathematics.  Our probabilistic projections relied on a novel combination of Bayesian growth curve modeling as well as methods for addressing model uncertainty via BMA, and relying on a large set of predictors derived from OECD, UNESCO, and World Bank data sources.  Our approach was based on obtaining an optimal prediction of the PISA 2009 starting minimum proficiency percentage and the rate of change in these percentages until 2022 and then plugging these estimates into a growth model to predict minimum proficiency percentage to 2033.


\par A benefit of BMA is that it provides a measure of variable importance on the basis of posterior inclusion probabilities. However, our results revealed that although one or two variables were deemed important across the reading and mathematics domains for boys and girls, their impact on the rate of progress in minimum proficiency was quite modest. This finding speaks to the need to develop parallel theory as to how changes in system level variables might impact change over time in minimum proficiency.  Developing such theory was beyond the scope of this paper.

\par A clear limitation of this work, and thus an opportunity for future research, concerns the sample of countries used in this analysis.  Specifically, for this study, we relied on data from OECD countries or those that partnered with the OECD to conduct PISA, as well as selecting those countries that provided data from 2009 to 2022.  Missing from this study are data from countries that did not participate in PISA. An important line of research going forward would be to harmonize data from ILSAs such as PISA with existing regional assessments such as the ERCE \citep{UNESCO_LLECE2020,UNESCOOREALC2023}, PASEC \citep{PASEC2019}, PILNA \citep{pilna2021}, SACMEQ \citep{SACMEQIV}, and SEA-PLM \citep{SEA-PLM2019}, to provide additional information not fully covered by the ILSAs, and indeed, countries making up these regional assessments are often (but not in all cases) those where noticeable inequities in literacy and numeracy are located.  This would be no small task insofar as these assessments have different goals, sampling designs, and assessment instruments. Regardless, even if harmonization across these international and regional assessments were successful, there are still many countries in the world with either very poor or non-existent data for monitoring minimum proficiency in reading and mathematics, making it extremely difficult to provide projections to 2030 \citep[see e.g.][]{Nilashi2023}.

\par With the above caveats and limitations in mind, we conclude that for the countries examined in this study, and with the novel approaches of combining growth models while addressing model uncertainty, all within a Bayesian perspective, that without interventions tailored to the needs of the individual countries, it seems unlikely that countries will reach their SDG 4.1.1c targets by 2030 and that the percentage of students at or above minimum proficiency in reading and mathematics will continue to remain stable or slowly decline as it has in the past. As a necessary driver for reaching many of the SDGs, quality education and providing equal learning opportunities for all must be an urgent focus of education policy efforts around the globe.

\backmatter

\newpage
\bibliographystyle{apacite}
\bibliography{sn-bibliography}

@book{berger2013statistical,
	title        = {Statistical Decision Theory and {B}ayesian Analysis},
	author       = {Berger, J.O.},
	year         = 2013,
	publisher    = {Springer}
}

@article{Breiman1996,
	title        = {Stacked regressions},
	author       = {Breiman, L},
	year         = 1996,
	journal      = {Machine Learning},
	volume       = 24,
	pages        = {49--64}
}

@incollection{clyde99,
	title        = {Bayesian model averaging and model search strategies (with discussion)},
	author       = {Clyde, M. A.},
	year         = 1999,
	booktitle    = {Bayesian Statistics, 6},
	publisher    = {Oxford University Press},
	address      = {Oxford},
	pages        = {157--185},
	editor       = {Bernardo, J. M. and Dawid, A. P. and Berger, J. O. and Smith, A. F. M.},
	editors      = {Berger, J. O., David, A. P., Smith, A. F. M.}
}

@article{clydegeorge,
	title        = {Model Uncertainty},
	author       = {Clyde, M. A. and George, E. I.},
	year         = 2004,
	journal      = {Statistical Science},
	volume       = 19,
	pages        = {81--94}
}

@incollection{ClydeIversen2013,
	title        = {Bayesian Model Averaging in the {M}-Open Framework},
	author       = {Clyde, M. A. and Iversen, E. S.},
	year         = 2013,
	booktitle    = {Bayesian Theory and Applications},
	publisher    = {Oxford University Press},
	address      = {Oxford},
	pages        = {483--498},
	editors      = {Damien P. and Dellaportas P. and Polson N. G. and Stephens D.A.}
}

@article{dawid84,
	title        = {Statistical theory: The prequential approach},
	author       = {Dawid, A. P.},
	year         = 1984,
	journal      = {Journal of the Royal Statistical Society, Series A},
	volume       = 147,
	pages        = {278--202}
}

@article{draper95,
	title        = {Assessment and propagation of model uncertainty (with discussion)},
	author       = {Draper, D.},
	year         = 1995,
	journal      = {Journal of the Royal Statistical Society (Series B)},
	volume       = 57,
	pages        = {55--98}
}

@article{eicher2011default,
	title        = {Default priors and predictive performance in {B}ayesian model averaging, with application to growth determinants},
	author       = {Eicher, T. S. and Papageorgiou, Chris and Raftery, Adrian E},
	year         = 2011,
	journal      = {Journal of Applied Econometrics},
	publisher    = {Wiley Online Library},
	volume       = 26,
	number       = 1,
	pages        = {30--55}
}

@book{feldkircher2009benchmark,
	title        = {Benchmark priors revisited: On adaptive shrinkage and the supermodel effect in {B}ayesian model averaging},
	author       = {Feldkircher, Martin and Zeugner, Stefan},
	year         = 2009,
	publisher    = {International Monetary Fund},
	number       = {9-202}
}

@article{fernandezleysteel01a,
	title        = {Benchmark priors for {B}ayesian model averaging},
	author       = {Fern\'{a}ndez, C. and Ley, E. and Steel, M. F. J.},
	year         = 2001,
	journal      = {Journal of Econometrics},
	volume       = 100,
	pages        = {381--427}
}

@article{Castillo2015,
  author    = {Castillo, I. and Schmidt-Hieber, J. and Van der Vaart, A.},
  title     = {Bayesian Linear Regression with Sparse Priors},
  journal   = {The Annals of Statistics},
  volume    = {43},
  number    = {5},
  pages     = {1986--2018},
  year      = {2015},
  doi       = {10.1214/15-AOS1334},
  mrnumber  = {3375874}
}

@article{fernandezleysteel01b,
	title        = {Model uncertainty in cross-country growth regressions},
	author       = {Fern\'{a}ndez, C. and Ley, E. and Steel, M. F. J.},
	year         = 2001,
	journal      = {Journal of Applied Econometrics},
	volume       = 16,
	pages        = {563--576}
}

@article{gneitingraftery,
	title        = {Strictly proper scoring rules, prediction, and estimation},
	author       = {Gneiting, T. and Raftery, A.},
	year         = 2007,
	journal      = {Journal of the American Statistical Association},
	volume       = 102,
	pages        = {359--378}
}

@article{Good1952,
	title        = {Rational Decisions},
	author       = {Good, I. J.},
	year         = 1952,
	journal      = {Journal of the Royal Statistical Society. Series B (Methodological)},
	volume       = 14,
	pages        = {107--114}
}

@article{KaplanHarraStampkaJude2025,
author={Kaplan, D. and Harra, K. and Stampka, J. and Jude, N.},
title={Stacking models of growth:
A methodology for predicting the pace of progress to the education
sustainable development targets using international large-scale assessments},
pages = {1-29},
journal={Psychometrika},
year={2025}}

@book{bernardosmith2000,
 author={Bernardo, J. and Smith, A. F. M.},
 title={Bayesian Theory},
 publisher={Wiley},
 address={New York},
 year={2000}}

@incollection{Akaike85,
   Author = {Akaike, H.},
   Title = {Prediction and entropy},
   BookTitle = {A Celebration of Statistics},
   Editor = {Atkinson, A. C. and Feinberg, S. E.},
   Publisher = {Springer-Verlag},
   Pages = {1-24},
      Year = {1985}}

@article{Akaike87,
   Author = {Akaike, H.},
   Title = {Factor analysis and {AIC}},
   Journal = {Psychometrika},
   Volume = {52},
   Pages = {317-332},
      Year = {1987}}

@article{nilashi2023,
  author    = {Nilashi, M. and Keng Boon, Ooi and Tan, G. and Lin, B. and Abumalloh, R.},
  title     = {Critical Data Challenges in Measuring the Performance of Sustainable Development Goals: Solutions and the Role of Big-Data Analytics},
  journal   = {Harvard Data Science Review},
  volume    = {5},
  year      = {2023},
  url       = {https://hdsr.mitpress.mit.edu/pub/9n4uzkg3},
  note      = {Accessed June 24, 2025}
}

@article{hoeting99,
	title        = {{B}ayesian Model Averaging: A Tutorial},
	author       = {Hoeting, J. A. and Madigan, D. and Raftery, A.  and Volinsky, C. T.},
	year         = 1999,
	journal      = {Statistical Science},
	volume       = 14,
	pages        = {382--417}
}

@article{JoseNauWinkler08,
	title        = {Scoring Rules, Generalized Entropy, and Utility Maximization},
	author       = {Jose, V. R. R. and Nau, R. F. and Winkler, R. L.},
	year         = 2008,
	journal      = {Operations Research},
	volume       = 56,
	pages        = {1146--1157}
}

@article{Kaplan2021Psychometrika,
	title        = {On the quantification of model uncertainty: A {B}ayesian perspective},
	author       = {Kaplan, David},
	year         = 2021,
	journal      = {Psychometrika},
	volume       = 86,
	pages        = {215--238},
}

@article{kaplanchenbma,
	title        = {Bayesian Model Averaging for Propensity Score Analysis},
	author       = {Kaplan, D. and Chen, J.},
	year         = 2014,
	journal      = {Multivariate Behavioral Research},
	volume       = 49,
	pages        = {505--517}
}

@article{KaplanHuang2021,
  author  = {Kaplan, David and Huang, Mingya},
  title   = {Bayesian probabilistic forecasting with large-scale educational trend data: a case study using NAEP},
  journal = {Large-scale Assessments in Education},
  year    = {2021},
  volume  = {9},
  pages   = {1-15},       
  doi     = {10.1186/s40536-021-00108-2},
  url     = {https://doi.org/10.1186/s40536-021-00108-2}
}

@article{kaplanleebmasem,
	title        = {{B}ayesian Model Averaging Over Directed Acyclic Graphs With Implications for the Predictive Performance of Structural Equation Models},
	author       = {Kaplan, D. and Lee, C},
	year         = 2015,
    pages          ={1-11},
	journal      = {Structural Equation Modeling},
}

@article{kassraftery,
	title        = {Bayes factors},
	author       = {Kass, R. E. and Raftery, A.},
	year         = 1995,
	journal      = {Journal of the American Statistical Association},
	volume       = 90,
	pages        = {773--795}
}

@article{KaplanLee2018,
	title        = {Optimizing prediction using {B}ayesian model averaging: Examples using large-scale educational assessments},
	author       = {Kaplan, D. and Lee, C.},
	year         = 2018,
	journal      = {Evaluation Review},
	doi          = {10.1177/0193841X18761421}
}

@misc{UNSDG,
	title        = {The 2030 Agenda for Sustainable Development},
	author       = {{UN General Assembly}},
	year         = 2015,
	note         = {Accessed 15 June 2019.},
	howpublished = {\url{https://www.refworld.org/docid/57b6e3e44.html}}
}

@manual{LaplacesDemon,
	title        = {LaplacesDemon: Complete Environment for Bayesian Inference},
	author       = {{Statisticat} and {LLC.}},
	year         = 2021,
	publisher    = {Bayesian-Inference.com},
	url          = {https://web.archive.org/web/20150206004624/http://www.bayesian-inference.com/software},
	note         = {R package version 16.1.6}
}

@book{leamer,
	title        = {Specification searches: Ad hoc inference with nonexperimental data},
	author       = {Leamer, E. E.},
	year         = 1978,
	publisher    = {Wiley},
	address      = {New York},
	volume       = 53
}

@article{LeySteel2009,
	title        = {On the effect of prior assumptions in Bayesian model averaging with applications to growth regression},
	author       = {Ley, E and Steel, M. F. J.},
	year         = 2009,
	journal      = {Journal of Applied Econometrics},
	volume       = 24,
	pages        = {651--674}
}

@book{lindley1991making,
	title        = {Making Decisions},
	author       = {Lindley, D. V.},
	year         = 1991,
	publisher    = {Wiley},
	address      = {London}
}

@misc{loo_R,
	title        = {loo: Efficient leave-one-out cross-validation and {WAIC} for {B}ayesian models},
	author       = {Aki Vehtari and Jonah Gabry and Yuling Yao and Andrew Gelman},
	year         = 2019,
	url          = {https://CRAN.R-project.org/package=loo},
	note         = {R package version 2.1.0}
}

@Manual{cran,
    title = {R: A Language and Environment for Statistical Computing},
    author = {{R Core Team}},
    organization = {R Foundation for Statistical Computing},
    address = {Vienna, Austria},
    year = {2023},
    url = {https://www.R-project.org/},
  }

@article{madiganraftery94,
	title        = {Model Selection and Accounting for Model Uncertainly in Graphical Models Using {O}ccam's Window},
	author       = {Madigan, D. and Raftery, A.},
	year         = 1994,
	journal      = {Journal of the American Statistical Association},
	volume       = 89,
	pages        = {1535--1546}
}

@book{GrimmRamEstabrook2017,
	title        = {Growth Modeling: Structural Equation and Multilevel Modeling Approaches},
	author       = {Grimm, K. J. and Ram, N. and Estabrook, R.},
	year         = 2017,
	publisher    = {Guilford},
	address      = {New York}
}

@article{mice,
	title        = {MICE: Multivariate Imputation by Chained Equations in {R}},
	author       = {{van} Buuren, S. and Groothuis-Oudshoorn, K.},
	year         = 2011,
    pages = {1-67},
	journal      = {Journal of Statistical Software}
}

@article{muthen1997latent,
  title={Latent variable modeling of longitudinal and multilevel data},
  author={Muth{\'e}n, Bengt},
  journal={Sociological Methodology},
  volume={27},
  pages={453--480},
  year={1997},
}

@article{raftery97,
	title        = {Bayesian model averaging for linear regression models},
	author       = {Raftery, A.  and Madigan, D. and Hoeting, J. A.},
	year         = 1997,
	journal      = {Journal of the American Statistical Association},
	volume       = 92,
	pages        = {179--191}
}

@article{rubin86fileconcat,
	title        = {Statistical Matching using File Concatenation with adjusted weights and multiple imputation},
	author       = {Rubin, D. B.},
	year         = 1986,
	journal      = {Journal of Business and Economic Statistics},
	volume       = 4,
	pages        = {87--95}
}

@article{sloughteretal,
	title        = {Probabilistic wind vector forecasting using ensembles and {B}ayesian model averaging},
	author       = {Sloughter, J. M. and Gneiting, T. and Raftery, A.},
	year         = {2013},
	journal      = {Monthly Weather Review},
	volume       = {141},
	pages        = {2107--2119}
}

@article{Tibshirani96,
	title        = {Regression Shrinkage and Selection via the Lasso},
	author       = {Tibshirani, R.},
	year         = 1996,
	journal      = {Journal of the Royal Statistical Society. Series B (Methodological)},
	volume       = 58,
	pages        = {267--288}
}

@article{vehtari2017loo,
	title        = {Practical {B}ayesian model evaluation using leave-one-out cross-validation and {WAIC}},
	author       = {Aki Vehtari and Andrew Gelman and Jonah Gabry},
	year         = 2017,
	journal      = {Statistics and Computing},
	volume       = 27,
	pages        = {1413--1432},
	doi          = {10.1007/s11222-016-9696-4},
	issue        = 5
}

@article{VehtariOjanen2012,
	title        = {A survey of {B}ayesian predictive methods for model assessment, selection and comparison},
	author       = {Vehtari, A. and Ojanen, J.},
	year         = 2012,
	journal      = {Statistics Surveys},
	publisher    = {Amer. Statist. Assoc., the Bernoulli Soc., the Inst. Math. Statist., and the~…},
	volume       = 6,
	pages        = {142--228},
	url          = {DOI: 10.1214/12-SS102}
}

@article{Watanabe2010,
	title        = {Asymptotic Equivalence of Bayes Cross Validation and Widely Applicable Information Criterion in Singular Learning Theory},
	author       = {Watanabe, S.},
	year         = 2010,
	journal      = {Journal of Machine Learning Research},
	volume       = 11,
	pages        = {3571--3594}
}

@article{willettsayer94,
	title        = {Using covariance structure analysis to detect correlates and predictors of individual change over time.},
	author       = {Willett, J. B. and Sayer, A. G.},
	year         = 1994,
	journal      = {Psychological Bulletin},
	volume       = 116,
	pages        = {363--381}
}

@article{Winkler96,
	title        = {Scoring rules and the evaluation of probabilities},
	author       = {Winkler, R. L.},
	year         = 1996,
	journal      = {Test},
	volume       = 5,
	pages        = {1--60}
}

@article{yeungbumgarnerraftery,
	title        = {Bayesian model averaging: development of an improved multi-class, gene selection, and classification tool for microarray data},
	author       = {Yeung, K. Y. and Bumgarner, R. E. and Raftery, A.},
	year         = {2005},
	journal      = {Bioinformatics},
	volume       = {21},
	pages        = {2394--2402}
}

@incollection{ZellnerGPrior,
	title        = {On Assessing Prior Distributions and {B}ayesian Regression Analysis with $g$ Prior Distributions},
	author       = {Zellner, A.},
	year         = 1986,
	booktitle    = {{B}ayesian Inference and Decision Techniques: Essays in Honor of {B}runo de {F}inetti. {S}tudies in {B}ayesian Econometrics},
	publisher    = {Elsevier},
	address      = {New York},
	pages        = {233--243},
	editor       = {Goel, P. and Zellner, A.}
}

@book{UNWomenUNDP2020,
  title        = {From Insights to Action: Gender Equality in the Wake of {COVID-19}},
  author       = {UNDP},
  year         = {2020},
    publisher={United Nations Development Program},
  institution  = {United Nations},
  address      = {New York, NY},
  url          = {https://www.unwomen.org/en/digital-library/publications/2020/09/gender-equality-in-the-wake-of-covid-19},
}

@book{pisa2000FirstResults,
	title        = {Knowledge and Skills for Life: First Results from {PISA} 2000},
	author       = {{OECD}},
	year         = 2001,
	publisher    = {Organization for Economic Cooperation and Development},
	address      = {Paris}
}

@techreport{SDGindex,
	title        = {{SDG} Index and Dashboards: Detailed Methodological paper},
	author       = {Guillaume Lafortune and Grayson Fuller and Jorge Moreno and Guido Schmidt-Traub and Christian Kroll},
	year         = 2018,
	note         = {\url{https://raw.githubusercontent.com/sdsna/2018GlobalIndex/master/\\2018GlobalIndexMethodology.pdf}},
  institution  = {Sustainable Development Solutions Network ({SDSN})},
}

@techreport{LevinB2003,
	title        = {Approaches to Equity in Policy for Lifelong Learning},
	author       = {Levin, B.},
	year         = {2003},
	note         = {A paper commissioned by the Education and Training Policy Division, OECD, for the Equity in Education Thematic Review},
	institution  = {OECD},
	howpublished = {\url{https://www.oecd.org/education/school/38692676.pdf}}
}

@misc{WorldBank2022LearningPoverty,
  author       = {{World Bank} and {UNICEF} and {UNESCO} and {FCDO} and {USAID} and {Bill \& Melinda Gates Foundation}},
  title        = {The State of Global Learning Poverty: 2022 Update},
  year         = {2022},
  url          = {https://www.unicef.org/media/122921/file/State%20of%20Learning%20Poverty%202022.pdf},
  note         = {Conference Edition, June 23, 2022},
  institution  = {World Bank Group}
}

@misc{vonDavier2024TIMSS,
  author       = {von Davier, Matthias and Kennedy, Adam and Reynolds, Kristin and Fishbein, Bethany and Khorramdel, Lale and Aldrich, Casey and Bookbinder, Ariel and Bezirhan, Umut and Yin, Liyang},
  title        = {TIMSS 2023 International Results in Mathematics and Science},
  year         = {2024},
  institution  = {Boston College, TIMSS \& PIRLS International Study Center},
  url          = {https://doi.org/10.6017/lse.tpisc.timss.rs6460},
  doi          = {10.6017/lse.tpisc.timss.rs6460}
}

@techreport{BorgonoviEtAl2017,
  author       = {Borgonovi, Francesca and Pokropek, Artur and Jakubowski, Maciej and Piacentini, Mario},
  title        = {Youth in Transition: How Do Some of the Cohorts Participating in PISA Fare in PIAAC?},
  institution  = {OECD Publishing},
  type         = {OECD Education Working Papers},
  number       = {155},
  address      = {Paris},
  year         = {2017},
  doi          = {10.1787/51479ec2-en},
  url          = {https://doi.org/10.1787/51479ec2-en}
}

@article{KullbackLeibler51,
author={Kullback, S. and Leibler, R. A.},
title={On information and sufficiency},
journal={Annals of Mathematical Statistics},
volume={22},
pages={79--86},
year={1951}}

@book{Kullback59,
author={Kullback,S.},
title={Information theory and statistics},
publisher={John Wiley and Sons},
address={New York},
year={1959}}

@article{Kullback87,
author={Kullback, S},
title={The {K}ullback-{L}eibler distance},
journal={The American Statistician},
volume={41},
pages={340--341},
year={1987}}

@book{OECD2008,
	title        = {Ten steps to equity in education. Policy brief},
	author       = {OECD},
	year         = {2008},
	publisher    = {OECD},
	address      = {Paris}
}

@incollection{Schleicher2008,
	title        = {Student Learning Outcomes in Mathematics from a Gender Perspective: What Does the International PISA Assessment Tell Us?},
	author       = {Schleicher, A.},
	year         = {2008},
	booktitle    = {Girls? Education in the 21st Century. Gender Equality, Empowerment, and Economic Growth},
	publisher    = {The World Bank},
	address      = {Washington},
	pages        = {41--52},
	editor       = {Tembon, M. and Fort, L.}
}

@book{EFAGMR2015,
	title        = {Education For All Global Monitoring Report 2015:  Achievements and Challenges},
	author       = {{UNESCO}},
	year         = {2015},
	publisher    = {{UNESCO}},
	address      = {Paris}
}

@book{UNESCO2016,
	title        = {Global Education Monitoring Report. Gender Review: Creating sustainable futures for all},
	author       = {UNESCO},
	year         = 2016,
	publisher    = {UNESCO},
	address      = {Paris}
}

@book{UIS2018,
	title        = {Handbook on Measuring Equity in Education},
	author       = {UNESCO},
	year         = 2018,
	publisher    = {UNESCO Institute for Statistics},
	address      = {Montreal}
}

@techreport{Montoya2022,
  author    = {Montoya, Silvia and MacDonald, Kevin},
  title     = {Monitoring of the Sustainable Development Goals using Large-Scale International Assessments: A strategy for reporting SDG 4 indicators using data from cross-national assessments},
  institution = {UNESCO Institute for Statistics},
  year      = {2022},
  url       = {https://tcg.uis.unesco.org/wp-content/uploads/sites/4/2022/04/Monitoring-of-the-SDGs-Using-Large-Scale-International-Assessments_April-2022.pdf}
}

@article{Kyriakides2017,
  author    = {Kyriakides, Leonidas and Georgiou, Marios P. and Creemers, Bert P. M. and Panayiotou, Anastasia and Reynolds, David},
  title     = {The impact of national educational policies on student achievement: a European study},
  journal   = {School Effectiveness and School Improvement},
  volume    = {29},
  number    = {2},
  pages     = {171--203},
  year      = {2017},
  doi       = {10.1080/09243453.2017.1398761},
  url       = {https://doi.org/10.1080/09243453.2017.1398761}
}

@incollection{Strietholt2019,
  author    = {Strietholt, R. and others},
  title     = {The Impact of Education Policies on Socioeconomic Inequality in Student Achievement: A Review of Comparative Studies},
  booktitle = {Socioeconomic Inequality and Student Outcomes},
  series    = {Education Policy \& Social Inequality},
  volume    = {4},
  editor    = {Volante, Louis and Schnepf, Sylke and Jerrim, John and Klinger, Don},
  publisher = {Springer},
  address   = {Singapore},
  year      = {2019},
  doi       = {10.1007/978-981-13-9863-6_2},
  url       = {https://doi.org/10.1007/978-981-13-9863-6\_2}
}

@techreport{pilna2021,
  title        = {{Pacific Islands Literacy and Numeracy Assessment ({PILNA}) 2021 Technical Report}},
  author       = {{Pacific Community}},
  year         = {2021},
  institution  = {Educational Quality and Assessment Programme (EQAP)},
  address      = {Suva, Fiji},
  publisher    = {{Pacific Community}},
  note         = {Available at  \url{https://pacificdata.org/data/dataset/pilna-2021-regional-report}}
}

@techreport{UNESCO_LLECE2020,
  author       = {UNESCO and LLECE},
  title        = {Análisis curricular del ERCE 2019 del conjunto de países que conforman la CECC/SICA},
  year         = {2020},
  institution  = {UNESCO},
  url          = {https://unesdoc.unesco.org/ark:/48223/pf0000375368}
}

@techreport{UNESCOOREALC2023,
  author       = {UNESCO-OREALC},
  title        = {Reporte técnico. Cuarto Estudio Regional Comparativo y Explicativo (ERCE 2019)},
  year         = {2023},
  institution  = {UNESCO-OREALC}
}

@techreport{SEA-PLM2019,
  author    = {UNICEF and Southeast Asian Ministers of Education Organization (SEAMEO) and Australian Council for Educational Research (ACER)},
  title     = {SEA-PLM 2019 Main Regional Report: Quality of Education in Southeast Asia},
  institution = {UNICEF, SEAMEO, ACER},
  year      = {2019},
  note = {\url{https://research.acer.edu.au/ar_misc/52/}}
}

@book{Encinas-Martin2023,
  author    = {Encinas-Martín, M. and Cherian, M.},
  title     = {Gender, Education and Skills: The Persistence of Gender Gaps in Education and Skills},
  series    = {OECD Skills Studies},
  publisher = {OECD Publishing},
  address   = {Paris},
  year      = {2023},
  doi       = {10.1787/34680dd5-en},
  url       = {https://doi.org/10.1787/34680dd5-en}
}

@misc{unesco2016education,
  title        = {Education 2030: Incheon Declaration and Framework for Action Towards Inclusive and Equitable Quality Education and Lifelong Learning for All},
  author       = {UNESCO},
  year         = {2016},
  howpublished = {\url{https://unesdoc.unesco.org/ark:/48223/pf0000245656}},
  note         = {Accessed: 2024-08-13}
}

@book{un2015mdg,
  title        = {The Millennium Development Goals Report 2015},
  author       = {{United Nations}},
  year         = {2015},
  publisher    = {United Nations Educational, Scientific and Cultural Organization (UNESCO)},
  address      = {New York, NY},
 howpublished={\url{https://www.un.org/millenniumgoals/2015_MDG_Report/pdf/MDG%202015%20rev%20(July%201).pdf}},
  
}

@incollection{Scheerens2023,
  author    = {Scheerens, Jaap},
  title     = {Theory on Teaching Effectiveness at Meta, General and Partial Level},
  booktitle = {Theorizing Teaching},
  editor    = {Praetorius, Anna-Katharina and Charalambous, Charalambos Y.},
  publisher = {Springer, Cham},
  year      = {2023},
  doi       = {10.1007/978-3-031-25613-4_4},
  url       = {https://doi.org/10.1007/978-3-031-25613-4\_4}
}

@article{kasswasserman,
author={Kass, R. E. and Wasserman, L.},
title={The selection of prior distributions by formal rules},
journal={Journal of the American Statistical Association},
volume={91},
pages={1343--1370},
year={1996}}

@Article{BMS,
    title = {Bayesian Model Averaging Employing Fixed and Flexible
      Priors: The {BMS} Package for {R}},
    author = {Stefan Zeugner and Martin Feldkircher},
    journal = {Journal of Statistical Software},
    year = {2015},
    volume = {68},
    number = {4},
    pages = {1--37},
    doi = {10.18637/jss.v068.i04},
  }

@techreport{PASEC2019,
  author    = {{CONFEMEN - Programme d'Analyse des Systèmes Éducatifs de la CONFEMEN}},
  title     = {{PASEC 2019 International Report: Quality of Education Systems in French-speaking Sub-Saharan Africa: Teaching/Learning Performance and Environment in Primary Education}},
  institution = {PASEC, CONFEMEN},
  year      = {2020},
  note = {\url{https://pasec.confemen.org/en/ressource/diagnostic-evaluation/}}
}

@techreport{SACMEQIV,
  author    = {Southern and Eastern Africa Consortium for Monitoring Educational Quality (SACMEQ)},
  title     = {SACMEQ IV: The Quality of Education in Southern and Eastern Africa},
  institution = {SACMEQ},
  year      = {2019},
  howpubished = {\url{https://www.seacmeq.org/?q=sacmeq-projects%2Fsacmeq-iv%2Freports}}
}

@article{MerkleSteyvers13,
author={Merkle, E. C. and Steyvers, M.},
title={Choosing a strictly proper scoring rule},
journal={Decision Analysis},
volume={10},
pages={292--304},
year={2013}}

@article{liang2008mixtures,
  title={Mixtures of g priors for Bayesian variable selection},
  author={Liang, Feng and Paulo, Rui and Molina, German and Clyde, Merlise A and Berger, Jim O},
  journal={Journal of the American Statistical Association},
  volume={103},
  pages={410--423},
  year={2008},
  publisher={Taylor \& Francis}
}

@article{Porwal2022,
  title={Comparing methods for statistical inference with model uncertainty},
  author={Porwal, Anupreet and Raftery, Adrian E.},
  journal={Proceedings of the National Academy of Sciences},
  volume={119},
  number={16},
  pages={e2120737119},
  year={2022},
  publisher={National Academy of Sciences},
  doi={10.1073/pnas.2120737119},
  url={https://www.pnas.org/doi/10.1073/pnas.2120737119}
}

@article{Porwal2023Effect,
  title={Effect of Model Space Priors on Statistical Inference with Model Uncertainty},
  author={Porwal, Anupreet and Raftery, Adrian E.},
  journal={New England Journal of Statistics in Data Science},
  volume={4},
  number={1},
  pages={1--20},
  year={2023},
  doi={10.51387/22-NEJSDS16},
  url={https://nejsds.nestat.org/journal/NEJSDS/article/16/read}
}

@article{Schwarz78,
   Author = {Schwarz, G.},
   Title = {Estimating the dimension of a model},
   Journal = {The Annals of Statistics},
   Volume = {6},
   Pages = {461-464},
   Year = {1978} }

@book{kaplanbayesbook2,
author={Kaplan, D},
title={Bayesian statistics for the social sciences},
edition={2nd},
publisher={Guilford Press},
address={New York},
year={2023}}

@techreport{OECD2018,
  title        = {Equity in Education: Breaking Down Barriers to Social Mobility},
  author       = {OECD},
  year         = {2018},
  institution  = {OECD},
  address      = {Paris},
  howpublished = {\url{https://doi.org/10.1787/9789264073234-en}
}}

@book{BollenCurran06,
	title        = {Latent Curve Models : A Structural Equation Perspective},
	author       = {Bollen, K. A. and Curran, P. J.},
	year         = 2006,
	publisher    = {John Wiley},
	address      = {New York}
}

@techreport{BorgonoviEtAl2018,
	title        = {The gender gap in educational outcomes in Norway},
	author       = {Borgonovi, F. and  Ferrara, A. and  Maghnouj, S.},
	year         = 2018,
	publisher    = {OECD},
	address      = {Paris},
	note         = {OECD Education Working Paper No. 183},
	institution  = {OECD}
}

@article{miBMA,
	title        = {An Approach to Addressing Multiple Imputation Model Uncertainty Using {B}ayesian Model Averaging},
	author       = {Kaplan, D. and Yavuz, S},
	year         = {2019},
	journal      = {Multivariate Behavioral Research},
	doi          = {10.1080/00273171.2019.1657790},
	url          = {https://doi.org/10.1080/00273171.2019.1657790},
	note         = {PMID: 31538505}
}

@book{littlerubin2019,
	title        = {Statistical analysis with missing data},
	author       = {Little, R. J. A. and Rubin, D. B},
	year         = 2019,
publisher = {John Wiley \& Sons},
	edition      = {3rd.},
}

@misc{naep,
  author       = {NCES},
  title        = {National Assessment of Educational Progress ({NAEP})},
  year         = {2022},
  howpublished = {\url{https://nces.ed.gov/nationsreportcard/}}
}

\end{document}